\RequirePackage{fix-cm} 
\documentclass[a4paper, twoside, reqno, dvips, 12pt]{amsart}
\usepackage{fixltx2e}   

\usepackage[latin1]{inputenc}
\usepackage[T1]{fontenc}

\usepackage{eucal}
\usepackage{esint}
\usepackage{dsfont}
\usepackage{xspace}
\usepackage{amsgen}
\usepackage{amsthm}
\usepackage{amssymb}
\usepackage{amsmath}
\usepackage{upgreek}
\usepackage{amsfonts}
\usepackage{MnSymbol}
\usepackage{mathrsfs}
\usepackage{mathtools}
\usepackage{textgreek}
\usepackage[nice]{nicefrac}

\usepackage{a4wide}

\usepackage[dvipsnames, table]{xcolor}
\definecolor{bckg}{RGB}{20.8, 20.8, 20.8}
\definecolor{oneblue}{rgb}{0.0, 0.0, 0.85}
\definecolor{Lightblue}{RGB}{214, 214, 214}
\definecolor{bluepigment}{rgb}{0.2, 0.2, 0.6}
\definecolor{charcoal}{rgb}{0.21, 0.27, 0.31}
\definecolor{denimblue}{rgb}{0.08, 0.38, 0.74}
\definecolor{Lightgray}{rgb}{0.89, 0.89, 0.89}
\definecolor{darkgrey}{rgb}{0.273, 0.281, 0.30}
\definecolor{darkelectricblue}{rgb}{0.33, 0.41, 0.47}

\usepackage[sort&compress, comma, square, numbers]{natbib}

\usepackage{psfrag}
\usepackage{graphicx}
\usepackage{subfigure}
\usepackage{morefloats}
\usepackage{indentfirst}
\usepackage[percent]{overpic}

\usepackage[perpage, symbol]{footmisc}

\usepackage{acronym}
\usepackage{microtype}
\usepackage[labelsep=period,%
            labelfont={bf,sf,color=bluepigment},%
            justification=raggedright]{caption}

\usepackage[usenames, dvipsnames, pdf]{pstricks}
\usepackage{epsfig}
\usepackage{pst-grad} 
\usepackage{pst-plot} 

\usepackage[colorlinks,
           urlcolor=oneblue,
           linkcolor=denimblue,
           citecolor=NavyBlue,
           bookmarksopen=false,
           pdfpagemode=UseNone,
           pagebackref]{hyperref}
           
\usepackage[explicit]{titlesec}

\titleformat{\section}
  {\color{NavyBlue}\Large\sffamily\bfseries}
  {}
  {0em}
  {\colorbox{bckg!5}{\parbox{\dimexpr\linewidth-2\fboxsep\relax}{\centering\thesection. #1}}}
  [\vspace*{0.33em}]

\titleformat{name=\section,numberless}
  {\color{NavyBlue}\Large\sffamily\bfseries}
  {}
  {0.0em}
  {\colorbox{bckg!10}{\parbox{\dimexpr\linewidth-2\fboxsep\relax}{\centering#1}}}
  [\vspace*{0.33em}]

\titleformat{\subsection}
  {\color{NavyBlue}\large\sffamily\bfseries}
  {}
  {0.0em}
  {\colorbox{bckg!5}{\parbox{\dimexpr\linewidth-2\fboxsep\relax}{\centering\thesubsection. #1}}}
  [\vspace*{0.33em}]

\titleformat{name=\subsection,numberless}
  {\color{NavyBlue}\Large\sffamily\bfseries}
  {}
  {0em}
  {\colorbox{bckg!10}{\parbox{\dimexpr\linewidth-2\fboxsep\relax}{\centering#1}}}
  [\vspace*{0.33em}]

\titleformat{\subsubsection}
  {\color{bluepigment}\sffamily\normalsize\bfseries}
  {\thesubsubsection}
  {0.5em}
  {#1}
  [\vspace*{0.33em}]

\titleformat{\paragraph}[runin]
  {\color{bluepigment}\sffamily\small\bfseries}
  {}
  {0em}
  {#1}

\titlespacing{\section}{1.0em}{1.5em plus 2pt minus 2pt}%
{1.0em plus 2pt minus 2pt}[0em]
\titlespacing{\subsection}{1.0em}{1.5em plus 2pt minus 2pt}%
{1.0em}[0em]
\titlespacing{\subsubsection}{1.0em}{1.5em plus 2pt minus 2pt}%
{1.0em plus 2pt minus 2pt}[0em]

\usepackage{titletoc}

\contentsmargin{0.5em}
\newlength{\tocsep} 
\titlecontents{section}[\tocsep]
  {\addvspace{10pt}\bfseries\sffamily}
  {\contentslabel[\thecontentslabel]{\tocsep}}
  {}
  {\ \titlerule*[0.75pc]{.}\ \thecontentspage}
  []
\titlecontents{subsection}[\tocsep]
  {\addvspace{8pt}\sffamily}
  {\contentslabel[\thecontentslabel]{\tocsep}}
  {}
  {\ \titlerule*[0.5pc]{.}\ \thecontentspage}
  []
\titlecontents*{subsubsection}[\tocsep]
  {\addvspace{2pt}\footnotesize\sffamily}
  {}
  {}
  {\ \titlerule*[0.35pc]{.}\ \thecontentspage}
  [\\*]

\makeatletter
\def\@setauthors{%
  \begingroup
  \def\thanks{\protect\thanks@warning}%
  \trivlist
  \centering\footnotesize \@topsep30\p@\relax
  \advance\@topsep by -\baselineskip
  \item\relax
  \author@andify\authors
  \def\\{\protect\linebreak}%
  \textsc{\normalsize\textcolor{darkelectricblue}{\authors}}%
  \ifx\@empty\contribs
  \else
    ,\penalty-3 \space \@setcontribs
    \@closetoccontribs
  \fi
  \endtrivlist
  \endgroup
}
\def\@settitle{\begin{center}%
  \baselineskip14\p@\relax
    \bfseries
    \textsc{\Large\textcolor{charcoal}{\@title}}
  \end{center}%
}
\makeatother

\usepackage{enumitem}
\setlist[description]{%
  topsep=30pt,               
  itemsep=5pt,               
  font={\bfseries\sffamily\color{NavyBlue}}, 
}

\usepackage{fancyhdr}
\usepackage{lastpage}

\newcommand*\Title{\textcolor{bluepigment}{Special solutions of the hyperbolic NLS equation}}
\newcommand*\Authors{\textcolor{bluepigment}{L.~Vuillon, D.~Dutykh \& F.~Fedele}}
\newcommand*{\plogo}{\textcolor{gray}{{\texttt{arXiv.org} / \textsc{hal}}}} 

\numberwithin{equation}{section}

\newcommand{\up}[1]{$^{\mathrm{\small\textsf{#1}}}$} 

\newcommand{\C}{\mathds{C}}
\newcommand{\R}{\mathds{R}}

\newcommand{\F}{\mathcal{F}}
\newcommand{\GZ}{\mathbb{Z}}
\newcommand{\A}{\mathcal{A}}

\newcommand{\N}{\mathcal{N}}

\newcommand{\eps}{\varepsilon}
\renewcommand{\O}{\mathcal{O}}
\renewcommand{\L}{\mathcal{L}}
\renewcommand{\H}{\mathcal{H}}
\renewcommand{\k}{\boldsymbol{k}}
\newcommand{\X}{\textrm{\textxi}}
\renewcommand{\S}{\ensuremath{\mathcal{S}}}

\newcommand{\Hh}{\mathbb{H}}
\newcommand{\Aa}{\mathbb{A}}
\newcommand{\Mm}{\mathbb{M}}

\newcommand{\ud}{\mathrm{d}}
\newcommand{\ue}{\mathrm{e}}
\newcommand{\ui}{\mathrm{i}}

\renewcommand{\Re}{\mathop{\mathrm{Re}}}
\renewcommand{\Im}{\mathop{\mathrm{Im}}}

\newcommand{\abs}[1]{\left|#1\right|}

\newcommand{\grad}{\boldsymbol{\nabla}}

\newcommand{\pd}[2]{\frac{\partial #1}{\partial\/ #2}}
\newcommand{\od}[2]{\frac{\mathrm{d} #1}{\mathrm{d}\/#2}}

\newcommand{\eqdef}{\mathop{\stackrel{\,\mathrm{def}}{:=}\,}}

\DeclareMathOperator{\spn}{span}

\newcommand{\cf}{\emph{cf.}\xspace}
\newcommand{\ie}{\emph{i.e.}\xspace}

\newcommand{\viz}{\emph{viz.}\xspace}
\newcommand{\etal}{\emph{et al.}\xspace}

\usepackage{acronym}
\acrodef{hNLS}[HypNLS]{Hyperbolic NLS}

\begin{document}

\title[\Title]{Some special solutions to the Hyperbolic NLS equation}

\author[L.~Vuillon]{Laurent Vuillon}
\address{\textbf{L.~Vuillon:} LAMA, UMR 5127 CNRS, Universit\'e Savoie Mont Blanc, Campus Scientifique, F-73376 Le Bourget-du-Lac Cedex, France}
\email{Laurent.Vuillon@univ-savoie.fr}
\urladdr{http://www.lama.univ-savoie.fr/~vuillon/}

\author[D.~Dutykh]{Denys Dutykh$^*$}
\address{\textbf{D.~Dutykh:} LAMA, UMR 5127 CNRS, Universit\'e Savoie Mont Blanc, Campus Scientifique, F-73376 Le Bourget-du-Lac Cedex, France}
\email{Denys.Dutykh@univ-savoie.fr}
\urladdr{http://www.denys-dutykh.com/}
\thanks{$^*$ Corresponding author}

\author[F.~Fedele]{Francesco Fedele}
\address{\textbf{F.~Fedele:} School of Civil and Environmental Engineering \& School of Electrical and Computer Engineering, Georgia Institute of Technology, Atlanta, USA}
\email{fedele@gatech.edu}
\urladdr{http://www.ce.gatech.edu/people/faculty/511/overview/}



\begin{titlepage}
\setcounter{page}{1}
\thispagestyle{empty} 
\noindent
{\Large Laurent \textsc{Vuillon}}\\
{\it\textcolor{gray}{CNRS--LAMA, University Savoie Mont Blanc, France}}\\[0.02\textheight]
{\Large Denys \textsc{Dutykh}}\\
{\it\textcolor{gray}{CNRS--LAMA, University Savoie Mont Blanc, France}}\\[0.02\textheight]
{\Large Francesco \textsc{Fedele}}\\
{\it\textcolor{gray}{Georgia Tech, Atlanta, USA}}\\[0.16\textheight]

\colorbox{Lightblue}{
  \parbox[t]{1.0\textwidth}{
    \centering\huge\sc
    \vspace*{0.7cm}

    \textcolor{bluepigment}{Some special solutions to the Hyperbolic NLS equation}

    \vspace*{0.7cm}
  }
}

\vfill 

\raggedleft     
{\large \plogo} 
\end{titlepage}


\newpage
\thispagestyle{empty} 
\par\vspace*{\fill}   
\begin{flushright} 
{\textcolor{denimblue}{\textsc{Last modified:}} \today}
\end{flushright}


\newpage
\maketitle
\thispagestyle{empty}


\begin{abstract}

The Hyperbolic Nonlinear \textsc{Schr\"odinger} equation (\acs{hNLS}) arises as a model for the dynamics of three--dimensional narrow-band deep water gravity waves. In this study, the symmetries and conservation laws of this equation are computed. The \textsc{Petviashvili} method is then exploited to numerically compute bi-periodic time-harmonic solutions of the \acs{hNLS} equation. In physical space they represent non-localized standing waves. Non-trivial spatial patterns are revealed and an attempt is made to describe them using symbolic dynamics and the language of substitutions. Finally, the dynamics of a slightly perturbed standing wave is numerically investigated by means a highly accurate \textsc{Fourier} solver.


\bigskip
\noindent \textbf{\keywordsname:} hyperbolic equations; NLS equation; wave patterns; deep water;  gravity waves; ground states \\


\end{abstract}


\newpage
\tableofcontents
\thispagestyle{empty}


\newpage
\section{Introduction}

The celebrated cubic Nonlinear \textsc{Schr\"odinger} (NLS) equation is one of the most important equations in nonlinear science \cite{Sulem1999}. For example, it arises in plasma physics \cite{Sen1978} and in normally dispersive optical waveguide arrays modeling \cite{Conti2003, Lahini2007}. In the context of water waves the \acs{hNLS} equation is the leading order model of the wave envelope evolution. It was derived for the first time by V.~\textsc{Zakharov} (1968) \cite{Zakharov1968} and rediscovered later by several other authors \cite{Chu1970, Hasimoto1972}. In one dimension ($1-$D), the NLS equation for unidirectional water waves is integrable \cite{Zakharov1972} and of focusing type. As a result, a steady or periodic balance of the cubic nonlinearities and wave dispersion can be attained, and this yields the formation of localized traveling waves (solitons) or homoclinic orbits to a plane wave (breathers). Analytical solutions for solitons follow via the inverse scattering transform \cite{Zakharov1972, John, Ablowitz1979, Pelinovsky2001a} and breathers can be easily obtained via the \textsc{Darboux} transformation \cite{Osborne2010}. The two-dimensional ($2-$D) propagation of deep water narrow-band waves is instead governed by the $2-$D \acl{hNLS} equation \cite{Ablowitz1979, Yuen1982, Sulem1999, Pelinovsky2003}. In this case, energy can spread along the transversal direction to the main propagation. A consequence of this de-focusing is that localized traveling waves cannot occur. Indeed, their non-existence was proved in (\cite{Ghidaglia1996a}, see also \cite{Ghidaglia1993}). However, this does not exclude the existence of nontrivial \emph{non-localized} travelling wave patterns that may arise due to a balance between nonlinearities and wave dispersion in both directions under toric constraints. To our knowledge there is still an open question of global well-posedness of solutions to the Initial Value Problem (IVP) for `large' initial data. Taking into account that there are very few analytical results for the hyperbolic NLS equation in contrast to its elliptic counterparts, the numerical approach adopted in our study is very appropriate.

An extended \acs{hNLS} equation was derived first by \textsc{Dysthe} (1979) \cite{Dysthe1979} and later by \textsc{Trulsen} \& \textsc{Dysthe} (1996) \cite{Trulsen1996}. Additionally to the classical hyperbolic (\textsc{D'Alembert}) operator, it contains also higher order dispersive and nonlinear terms. The generalization to the finite depth case leads to the \textsc{Davey}--\textsc{Stewartson} (DS) equations \cite{Benney1967, Davey1974, Ghidaglia1990}. Several important analytical solutions to the the DS model were derived in \cite{Ebadi2011, Ebadi2011a, Yildirim2012, Jafari2012}.

For mathematical/numerical studies it is often convenient to restrict the attention to a particular class of solutions. For example, the very first mathematical description of plane permanent waves is known at least since G.~\textsc{Stokes} (1847) \cite{Stokes1847}. Permanent waves can be periodic or localized in space. In this study we focus on bi-periodic (in spatial dimensions) time-harmonic solutions of the form $A\,(x,\,y,\,t)\ \equiv\ F\,(x\ -\ c_{\,g}\, t,\, y)\;\ue^{\ui\,\omega\,t}\,$, where $A:\ \R^{\,2}\times \R^{\,+}\ \mapsto\ \C$ is the complex envelope and $F:\ \R^2\ \mapsto\ \C$ a periodic complex function with respect to its two arguments. These are stationary solutions in the frame of reference moving with the group speed $c_{\,g}\ \in\ \R\,$. In the physical domain, the associated wave surface displacements is that of standing waves, which have been the subject of many studies. In particular, radial standing solutions of the \acs{hNLS} equation were investigated in \cite{Kevrekidis2011}. The existence of standing waves in deep waters was proved in \cite{Iooss2005}. \textsc{Sulem} \& \textsc{Sulem} (1999) report the results of numerical simulations of the \acs{hNLS} equation in (bi-)periodic domains \cite{Sulem1999}. In contrast to the (elliptic) $1-$D and $2-$D focusing cases, no tendency to collapse was observed numerically. On the other hand, an approximate \textsc{Fermi}--\textsc{Pasta}--\textsc{Ulam}(--\textsc{Tsingou}) (FPU) recurrence \cite{Fermi1955} was reported. \textsc{Osborne} \etal (2000) investigated the existence of coherent structures in the \acs{hNLS} equations \cite{Osborne2000}, since their very existence has been left in doubt. They performed numerical experiments and concluded that unstable modes exist in the \acs{hNLS} equation dynamics in $2-$D.

In the shallow-water regime standing wave patterns have been found in the context of the \textsc{Boussinesq} equations by M.~\textsc{Chen} \& G.~\textsc{Iooss} \cite{Chen2005, Chen2006, Chen2008}. $2-$D bi-periodic travelling wave solutions to the \textsc{Euler} equations were studied by W.~\textsc{Craig} \& D.~\textsc{Nicholls} (2002) \cite{Craig2002}. Recently, $2-$D wave patterns of the free surface were investigated experimentally by D.~\textsc{Henderson} \etal (2010) \cite{Henderson2010} along with a theoretical stability analysis.

In this work, symbolic dynamics and associated techniques of substitutions are exploited to investigate the structure of standing wave patterns (see \cite{Morse1940, Lind1995, Berthe2000}). In physics, their application in studies of dynamical systems led to unveiling the structure of quasi-periodic tilings (see \cite{Senechal1995}). Symbolic dynamics allows coding the nonlinear behavior of a complex system and pattern formation by means of $1-$D or $2-$D words of a finite alphabet. Such approach was applied to code the non-periodic trajectories on a unit circle with a particular partition on two intervals (see \cite{Morse1940}). Coding of finite, periodic and non-periodic infinite patterns using $2-$D words and tilings was done in \cite{Berthe2000, DeBruijn1981}. The numerical standing waves investigated in this work are periodic in space, thus the associated patterns are described up to toric constraints. This is the first step for developing a theory for the description of periodic or quasi-periodic patterns associated to trajectories $\{a_{\,n}\,(t)\}_{\,n}$ of the \acs{hNLS} dynamics in the infinite phase space spanned by, for example, generalized \textsc{Fourier} basis $\phi_{\,n}\,(x,\,y)$ associated to the formal series for $A\,(x,\,y,\,t)\ \simeq\ \sum_n\; a_{\,n}\,(t)\cdot \phi_{\,n}\,(x,\,y)$ on a periodic domain. The evidences presented below suggest that efficient reduced-order descriptions are possible for the \acs{hNLS} equation along the lines of \cite{Lucia2004}, to cite an example.

The present study is organized as follows. First, the \acs{hNLS} equation is introduced in the context of deep water waves. Then, the \textsc{Petviashvili} method \cite{Petviashvili1976} used to compute a class of special standing wave solutions is presented. In the spirit of \textsc{Sturm}'s works, the symbolic dynamics is then applied to describe the associated spatial patterns using words and substitutions. This idea stems from the pioneering work of \textsc{Sturm} (1836), who used the \emph{coding} of trajectories to describe the solutions to PDEs \cite{Sturm1836}. See also the book \cite{Galaktionov2004} for a modern account of this theory. This coding is referred to as `symbolic dynamics' since the work of \textsc{Morse} \& \textsc{Hedlund} (1940) \cite{Morse1940}. In this work, symbolic dynamics and associated techniques of substitutions are exploited to investigate the structure of standing wave patterns arising in the \acs{hNLS} equation. Finally, a highly accurate \textsc{Fourier}-based solver is exploited to investigate the dynamics of a perturbed standing wave.


\section{The Mathematical Model}
\label{sec:math}

Consider a three-dimensional fluid domain with a free surface. The water is assumed to be infinitely deep. The \textsc{Cartesian} coordinate system $O\,x\,y\,z$ is chosen such that the undisturbed water level corresponds to $z\ =\ 0\,$, and the free surface elevation is $z\ =\ \eta\,(x,\,y,\,t)\,$.

The Euler equations that describe the irrotational flow of an ideal incompressible fluid of infinite depth with a free surface are of fundamental relevance in fluid mechanics, ocean sciences and both pure and applied mathematics (see for example, \cite{Stoker1957, Zakharov1968, Johnson1997}). The structure of the \textsc{Euler} equations is given in terms of the free-surface elevation $\eta\,(x,\, y,\, t)$ and the velocity potential $\varphi\,(x,\, y,\, t)\ =\ \phi\,(x,\, y,\, z\ =\ \eta\,(x,\,y,\,t),\,t)$ evaluated at the free surface of the fluid. In late 70's \textsc{Dysthe} used the method of multiple scales to derive from the \textsc{Euler} equations a modified Nonlinear \textsc{Schr\"odinger} (NLS) equation \cite{Dysthe1979} for the time evolution of the unidirectional narrow-band envelope $A$ of the velocity potential $\varphi$ with carrier wave $\ue^{\,\ui\,(k_{\,0}\,x\ -\ \omega_{\,0}\,t)}\,$, $k_{\,0}$ and $\omega_{\,0}$ being, respectively, the wavenumber and frequency of the carrier wave. The equation for $A$ can be formulated in a frame moving with the group velocity $c_{\,g}\ =\ \frac{\omega_{\,0}}{2\,k_{\,0}}$ as follows. Define $\eps$ as a small parameter, $a_{\,0}$ as a characteristic wave amplitude and rescale space, time and the envelope as  $x\ \to\ k_{\,0}\,x\ -\ c_{\,g}\,t\,$, $y\ \to\ k_{\,0}\,y\,$, $t\ \to\ \omega_{\,0}\,t$ and $A\ \to\ \eps\,a_{\,0}\,A$ respectively. Then, the $2-$D \textsc{Dysthe} equation for $A$ is given by \cite{Trulsen1996}:
\begin{multline}\label{eq:Dysthe}
  \ui\, A_{\,t}\ =\ \frac{1}{8}\;A_{\,x\,x}\ -\ \frac{1}{4}\;A_{\,y\,y}\ +\ \frac{1}{2}\;\abs{A}^{\,2}\,A \\ -\ \ui\,\eps\,\Bigl(\frac{1}{16}\;A_{\,x\,x\,x}\ -\ \frac{3}{8}\;A_{\,x\,y\,y}\ -\ \frac{3}{2}\,\abs{A}^{\,2}\,A_{\,x}\ +\ \frac{1}{4}\;A^{\,2}\,A^{\,\star}_{\,x}\ -\ \ui\,A\,\H[\abs{A}^{\,2}]\Bigr)\,,
\end{multline}
where $\H\,[\cdot]$ is the \textsc{Hilbert} transform and the subscripts $A_{\,t}\ =\ \partial_{\,t}\,A\,$, $A_{\,x}\ =\ \partial_{\,x}\,A$ denote partial derivatives with respect to $x\,$, $y$ and $t$ respectively, and $A^{\,\star}$ denotes complex conjugation. The envelope of the free surface $\eta$ relates to $A$ by a simple transformation that involves only $A$ and its derivatives (see \cite{Dysthe1979}). To $\O\,(1)$ in $\eps\,$, the $2-$D \textsc{Dysthe} equation \eqref{eq:Dysthe} reduces to the \acs{hNLS} equation:
\begin{equation}\label{eq:hNLS}
  \ui\, A_{\,t}\ =\ \frac{1}{8}\;A_{\,x\,x}\ -\ \frac{1}{4}\;A_{\,y\,y}\ +\ \frac{1}{2}\;\abs{A}^{\,2}\,A\,.
\end{equation}
By rescaling $A\ \to\ 2\,A\,$, $x\ \to\ 2\,\sqrt{2}\, x$ and the transverse coordinate $y\ \to\ 2\,y\,$, \eqref{eq:hNLS} takes the form \cite{Sulem1999, Pelinovsky2003}:
\begin{equation}\label{eq:adim}
  \ui\,A_{\,t}\ =\ A_{\,x\,x}\ -\ A_{\,y\,y}\ +\ 2\,A\,\abs{A}^{\,2}\,.
\end{equation}
This equation admits the three invariants $\Hh\,$, $\Aa$ and $\Mm\,$, which have the meaning of energy, wave action and momentum respectively:
\begin{eqnarray*}
  \Hh\ &=&\ \iint_{\;\R^{\,2}}\Bigl\{\;\abs{A_{\,y}}^{\,2}\ -\ \abs{A_{\,x}}^{\,2}\ +\ \abs{A}^{\,4}\;\Bigr\}\,\ud x\,\ud y\,, \\
  \Aa\ &=&\ \iint_{\;\R^{\,2}}\;\abs{A}^{\,2}\,\ud x\,\ud y\,, \\
  \Mm\ &=&\ \frac{\ui}{2}\;\iint_{\;\R^{\,2}}\Bigl\{\;A\,\grad A^{\,\star}\ -\ A^{\,\star}\,\grad A\;\Bigr\}\,\ud x\,\ud y\,.
\end{eqnarray*}
Note that the total energy $\Hh$ is also the \textsc{Hamiltonian} for the \acs{hNLS} equation. In the following, we will solve for a special class of standing wave solutions to \eqref{eq:adim}.

For the sake of convenience, equation \eqref{eq:adim} can be equivalently rewritten as a system of two coupled PDEs for two real-valued functions. By $u\,(x,\,y,\,t)\ \eqdef\ \Re A\,(x,\,y,\,t)$ and $v\,(x,\,y,\,t)\ \eqdef\ \Im A\,(x,\,y,\,t)$ we denote correspondingly real and imaginary parts of the complex wave envelope $A\,(x,\,y,\,t)\,$. The resulting system of PDEs then reads:
\begin{align}\label{eq:adim1}
  u_{\,t}\ &=\ v_{\,x\,x}\ -\ v_{\,y\,y}\ +\ 2\,v\,(u^{\,2}\ +\ v^{\,2})\,, \\
  v_{\,t}\ &=\ -\,u_{\,x\,x}\ +\ u_{\,y\,y}\ -\ 2\,u\,(u^{\,2}\ +\ v^{\,2})\,.\label{eq:adim2}
\end{align}


\subsection{Symmetries of the evolution equation}

The \textsc{Lie} algebra of infinitesimal (or symmetry) generators of the system of real-valued governing equations \eqref{eq:adim1}, \eqref{eq:adim2} can be readily computed \cite{Cheviakov2007}:
\begin{equation*}
  \X_{\,1}\ =\ \pd{}{t}\,, \quad
  \X_{\,2}\ =\ \pd{}{x}\,, \quad
  \X_{\,3}\ =\ \pd{}{y}\,, \quad
  \X_{\,4}\ =\ x\;\pd{}{y}\ +\ y\;\pd{}{x}\,,
\end{equation*}
\begin{equation*}
  \X_{\,5}\ =\ u\;\pd{}{v}\ -\ v\;\pd{}{u}\,, \quad
  \X_{\,6}\ =\ \frac{1}{2}\;x\,v\;\pd{}{u}\ -\ \frac{1}{2}\;u\,x\,\pd{}{v}\ +\ t\;\pd{}{x}
\end{equation*}
\begin{equation*}
  \X_{\,7}\ =\ -\,\frac{1}{2}\;y\,v\;\pd{}{u}\ +\ \frac{1}{2}\;u\,y\,\pd{}{v}\ +\ t\;\pd{}{y}\,, \quad
  \X_{\,8}\ =\ 2\,t\;\pd{}{t}\ -\ u\;\pd{}{u}\ -\ v\;\pd{}{v}\ +\ x\;\pd{}{x}\ +\ y\;\pd{}{y}\,,
\end{equation*}
\begin{equation*}
  \X_{\,9}\ =\ \biggl[\,\frac{1}{4}\;\bigl(x^{\,2}\ -\ y^{\,2}\,\bigr)\,v\ -\ t\,u\,\biggr]\;\pd{}{u}\ -\ \biggl[\,\frac{1}{4}\;\bigl(x^{\,2}\ -\ y^{\,2}\bigr)\,u\ +\ t\,v\,\biggr]\;\pd{}{v}\ +\ t^{\,2}\;\pd{}{t}\ +\ t\,x\;\pd{}{x}\ +\ t\,y\;\pd{}{y}\,.
\end{equation*}

Point symmetries of the PDE \eqref{eq:adim} can be obtained from infinitesimal generators $\bigl\{\X_{\,i}\bigr\}_{i\,=\,1}^{\,9}$ listed hereinabove by solving induced systems of differential equations. The corresponding transformations are given below:
\begin{equation*}
  (1)\ \quad t^{\,\prime}\ =\ t\ +\ \eps_{\,1}\,, \quad x^{\,\prime}\ =\ x\,, \quad y^{\,\prime}\ =\ y\,, \quad u^{\,\prime}\ =\ u\,, \quad v^{\,\prime}\ =\ v\,,
\end{equation*}
\begin{equation*}
  (2)\ \quad t^{\,\prime}\ =\ t\,, \quad x^{\,\prime}\ =\ x\ +\ \eps_{\,2}\,, \quad y^{\,\prime}\ =\ y\,, \quad u^{\,\prime}\ =\ u\,, \quad v^{\,\prime}\ =\ v\,,
\end{equation*}
\begin{equation*}
  (3)\ \quad t^{\,\prime}\ =\ t\,, \quad x^{\,\prime}\ =\ x\,, \quad y^{\,\prime}\ =\ y\ +\ \eps_{\,3}\,, \quad u^{\,\prime}\ =\ u\,, \quad v^{\,\prime}\ =\ v\,,
\end{equation*}
\begin{multline*}
  (4)\ \quad t^{\,\prime}\ =\ t\,, \quad x^{\,\prime}\ =\ \frac{x\ +\ y}{2}\;\ue^{\,\eps_{\,4}}\ +\ \frac{x\ -\ y}{2}\;\ue^{\,-\eps_{\,4}}\,, \\
  y^{\,\prime}\ =\ \frac{x\ +\ y}{2}\;\ue^{\,\eps_{\,4}}\ +\ \frac{y\ -\ x}{2}\;\ue^{\,-\eps_{\,4}}\,, \quad u^{\,\prime}\ =\ u\,, \quad v^{\,\prime}\ =\ v\,,
\end{multline*}
\begin{multline*}
  (5)\ \quad t^{\,\prime}\ =\ t\,, \quad x^{\,\prime}\ =\ x\,, \quad y^{\,\prime}\ =\ y\,, \\
  u^{\,\prime}\ =\ \cos(\eps_{\,5})\,u\ -\ \sin(\eps_{\,5})\,v\,, \quad
  v^{\,\prime}\ =\ \sin(\eps_{\,5})\,u\ +\ \cos(\eps_{\,5})\,v\,,
\end{multline*}
\begin{multline*}
  (6)\ \quad t^{\,\prime}\ =\ t\,, \quad x^{\,\prime}\ =\ x\ +\ \eps_{\,6}\,t\,, \quad y^{\,\prime}\ =\ y\,, \\
  u^{\,\prime}\ =\ \cos\,\Bigl(\frac{\eps_{\,6}}{2}\;\bigl(x\ +\ \frac{\eps_{\,6}}{2}\;t\bigr)\Bigr)\,u\ +\ \sin\,\Bigl(\frac{\eps_{\,6}}{2}\;\bigl(x\ +\ \frac{\eps_{\,6}}{2}\;t\bigr)\Bigr)\,v\,, \\
  v^{\,\prime}\ =\ -\sin\,\Bigl(\frac{\eps_{\,6}}{2}\;\bigl(x\ +\ \frac{\eps_{\,6}}{2}\;t\bigr)\Bigr)\,u\ +\ \cos\,\Bigl(\frac{\eps_{\,6}}{2}\;\bigl(x\ +\ \frac{\eps_{\,6}}{2}\;t\bigr)\Bigr)\,v\,,
\end{multline*}
\begin{multline*}
  (7)\ \quad t^{\,\prime}\ =\ t\,, \quad x^{\,\prime}\ =\ x\,, \quad y^{\,\prime}\ =\ y\ +\ \eps_{\,7}\,t\,, \\
  u^{\,\prime}\ =\ \cos\,\Bigl(\frac{\eps_{\,7}}{2}\;\bigl(y\ +\ \frac{\eps_{\,7}}{2}\;t\bigr)\Bigr)\,u\ -\ \sin\,\Bigl(\frac{\eps_{\,7}}{2}\;\bigl(y\ +\ \frac{\eps_{\,7}}{2}\;t\bigr)\Bigr)\,v\,, \\
  v^{\,\prime}\ =\ \sin\,\Bigl(\frac{\eps_{\,7}}{2}\;\bigl(y\ +\ \frac{\eps_{\,7}}{2}\;t\bigr)\Bigr)\,u\ +\ \cos\,\Bigl(\frac{\eps_{\,7}}{2}\;\bigl(y\ +\ \frac{\eps_{\,7}}{2}\;t\bigr)\Bigr)\,v\,,
\end{multline*}
\begin{equation*}
  (8)\ \quad t^{\,\prime}\ =\ \ue^{\,2\,\eps_{\,8}}\,t\,, \quad
  x^{\,\prime}\ =\ \ue^{\,\eps_{\,8}}\,x\,, \quad
  y^{\,\prime}\ =\ \ue^{\,\eps_{\,8}}\,y\,, \quad
  u^{\,\prime}\ =\ \ue^{\,-\eps_{\,8}}\,u\,, \quad
  v^{\,\prime}\ =\ \ue^{\,-\eps_{\,8}}\,v\,,
\end{equation*}
\begin{multline*}
  (9)\ \quad t^{\,\prime}\ =\ \frac{t}{1\ -\ \eps_{\,9}\,t}\,, \quad
  x^{\,\prime}\ =\ \frac{x}{1\ -\ \eps_{\,9}\,t}\,, \quad
  y^{\,\prime}\ =\ \frac{y}{1\ -\ \eps_{\,9}\,t}\,, \\
  u^{\,\prime}\ =\ (1\ -\ \eps_{\,9}\,t)\,\biggl[\,\cos\biggl(\frac{\eps_{\,9}\,\bigl(x^{\,2}\ -\ y^{\,2}\bigr)}{4\,(1\ -\ \eps_{\,9}\,t)}\biggr)\,u\ +\ \sin\biggl(\frac{\eps_{\,9}\,\bigl(x^{\,2}\ -\ y^{\,2}\bigr)}{4\,(1\ -\ \eps_{\,9}\,t)}\Bigr)\,v\,\biggr]\,, \\
  v^{\,\prime}\ =\ (1\ -\ \eps_{\,9}\,t)\,\biggl[\,-\sin\biggl(\frac{\eps_{\,9}\,\bigl(x^{\,2}\ -\ y^{\,2}\bigr)}{4\,(1\ -\ \eps_{\,9}\,t)}\biggr)\,u\ +\ \cos\biggl(\frac{\eps_{\,9}\,\bigl(x^{\,2}\ -\ y^{\,2}\bigr)}{4\,(1\ -\ \eps_{\,9}\,t)}\biggr)\,v\,\biggr]\,,
\end{multline*}
where $\{\eps_{\,i}\}_{\,i\,=\,1}^{\,9}\ \in\ \R$ are transformations parameters and they do not have to be small. The `geometrical' sense of these transformations is the following:
\begin{description}
  \item[(1)] time translation
  \item[(2,3)] spatial translations
  \item[(4)] hyperbolic plane transformation (see Section~\ref{sec:patt} for explanations)
  \item[(5)] phase translation (in the complex case) corresponds to the solution rotation in the real case
  \item[(6,7)] \textsc{Galilean} boosts along coordinate axes
  \item[(8)] scaling transformation
  \item[(9)] a genuinely local transformation whose physical/geometrical sense is not clear
\end{description}


\subsection{Standing wave patterns}
\label{sec:gs}

Consider the ansatz for standing wave solutions that oscillate harmonically in time
\begin{equation}\label{eq:ansatz}
  A\,(x,\,y,\,t)\ =\ \ue^{\,-\ui\,\omega\,t}\;B\,(x,\,y)\,,
\end{equation}
where the real function $B\,(x,\,y)\;:\ \R^{\,2}\ \mapsto\ \R$ describes the spatial pattern of a standing wave. According to the unscaled \acs{hNLS} equation \eqref{eq:adim}, $B\,(x,\,y)$ satisfies the following real nonlinear hyperbolic PDE
\begin{equation}\label{eq:Beq}
  \omega\,B\ +\ B_{\,x\,x}\ -\ B_{\,y\,y}\ =\ 2\,B^{\,3}\,,
\end{equation}


\subsubsection{Symmetries of standing wave patterns}
\label{sec:patt}

The symmetry group of point transformations preserving solutions to equation \eqref{eq:Beq} can be also computed. The infinitesimal generators are given here:
\begin{equation*}
  \X_{\,1}\ =\ \pd{}{x}\,, \qquad
  \X_{\,2}\ =\ \pd{}{y}\,, \qquad
  \X_{\,3}\ =\ y\;\pd{}{x}\ +\ x\;\pd{}{y}\,.
\end{equation*}
One can see that the generated \textsc{Lie} algebra is much smaller comparing to the evolution equation. Thus, standing wave patterns represent a more `rigid' coherent structure. It translates the fact that we are considering special solutions satisfying ansatz \eqref{eq:ansatz}. The corresponding point transformations can be readily computed:
\begin{equation*}
  (1)\ \quad x^{\,\prime}\ =\ x\ +\ \eps_{\,1}\,, \quad y^{\,\prime}\ =\ y\,, \quad B^{\,\prime}\ =\ B\,,
\end{equation*}
\begin{equation*}
  (2)\ \quad x^{\,\prime}\ =\ x\,, \quad y^{\,\prime}\ =\ y\ +\ \eps_{\,2}\,, \quad B^{\,\prime}\ =\ B\,,
\end{equation*}
\begin{equation*}
  (3)\ \quad x^{\,\prime}\ =\ \frac{x\ +\ y}{2}\;\ue^{\,\eps_{\,3}}\ +\ \frac{x\ -\ y}{2}\;\ue^{\,-\eps_{\,3}}\,, \quad
  y^{\,\prime}\ =\ \frac{x\ +\ y}{2}\;\ue^{\,\eps_{\,3}}\ +\ \frac{y\ -\ x}{2}\;\ue^{\,-\eps_{\,3}}\,, \quad
  B^{\,\prime}\ =\ B\,,
\end{equation*}
where $\{\eps_{\,i}\}_{\,i\,=\,1}^{\,3}\ \in\ \R$ are transformation parameters. Transformations (1,2) are shifts along coordinate axes. In order to understand geometrical meaning of transformation (3), we rewrite it in an equivalent form:
\begin{equation*}
  (3^{\,\prime}) \quad x^{\,\prime}\ =\ \cosh(\eps_{\,3})\,x\ +\ \sinh(\eps_{\,3})\,y\,, \qquad
  y^{\,\prime}\ =\ \sinh(\eps_{\,3})\,x\ +\ \cosh(\eps_{\,3})\,y\,,
\end{equation*}
where we dropped out the dependent ($u\,$, $v$) and independent ($t$) variables, which are not affected by this transformation. Now it is clear that $(3^{\,\prime})$ is a hyperbolic transformation of the plane, which preserves the orientation and the pseudo-\textsc{Euclidean} metrics, \ie
\begin{equation*}
  (x^{\,\prime})^{\,2}\ -\ (y^{\,\prime})^{\,2}\ \equiv\ x^{\,2}\ -\ y^{\,2}\,.
\end{equation*}


\subsubsection{Numerical illustrations}

Equation~\eqref{eq:Beq} can be solved numerically using the classical \textsc{Petviashvili} method \cite{Petviashvili1976, Pelinovsky2004, Lakoba2007}. To do so, \eqref{eq:Beq}  is rewritten in the operator form:
\begin{equation*}
  \L\cdot B\ =\ \N\,(B), \qquad \L\ \eqdef\ \omega\ +\ \partial_{\,x\,x}\ -\ \partial_{\,y\,y}\,, \qquad \N\,(B) \eqdef 2\,B^{\,3}\,.
\end{equation*}
where $\S$ is the so-called stabilizing factor\footnote{The purpose of this stabilizing factor is to remove the unstable eigenvalue of the iteration matrix. This technology is well understood today, \cf \cite{Alvarez2014}.} and the exponent $\gamma$ is usually defined as a function of the degree of nonlinearity $p$ ($p\ =\ 3$ for the \acs{hNLS} equation). The rule of thumb prescribes the following formula $\gamma\ =\ \dfrac{p}{p\ -\ 1}\,$. The scalar product is defined in the $L_{\,2}$ space. The inverse operator $\L^{\,-1}$ can be efficiently computed in \textsc{Fourier} space using the Fast \textsc{Fourier} Transform (FFT) (see, for example, \cite{Frigo2005}), and spatial periodicity across the 2D computational box boundaries is implicitly imposed. The iterative process converges for a large class of smooth initial guesses\footnote{In our computations we just took an initial localized bump in the center of the domain and the iterative method turned it into a pattern, if the domain $\Omega$ size and the frequency $\omega$ were chosen accordingly.}. Convergence is attained when the $L_{\,\infty}$ norm between two successive iterations is less than a prescribed tolerance $\varepsilon$ (usually of the order of machine precision). Additionally, the residual error $E_{\,r}$ in approximating the nonlinear equation is checked by substituting in \eqref{eq:Beq} the converged numerical solution. In the present work, the convergence of the algorithm is checked numerically in the extended floating point arithmetics using $30$ significant digits \cite{MATLAB2012}. We reiterate that the resulting fully-converged pattern depends only on the parameter $\omega$ and the computational domain $\Omega\,$. All the patterns presented below are generated in the following way. First, we fix the computational domain $\Omega$ based on a number-theoretical reasoning. Then, we make a search over the parameter $\omega\ \in\ \bigl(0,\,\omega_{\,\max}\,\bigr]$ to pick up the \emph{discrete} values for which the convergence is achieved, which is equivalent here to the emergence of periodic patterns.

\begin{figure}
  \centering
  \subfigure[]{%
  \includegraphics[width=0.70\textwidth]{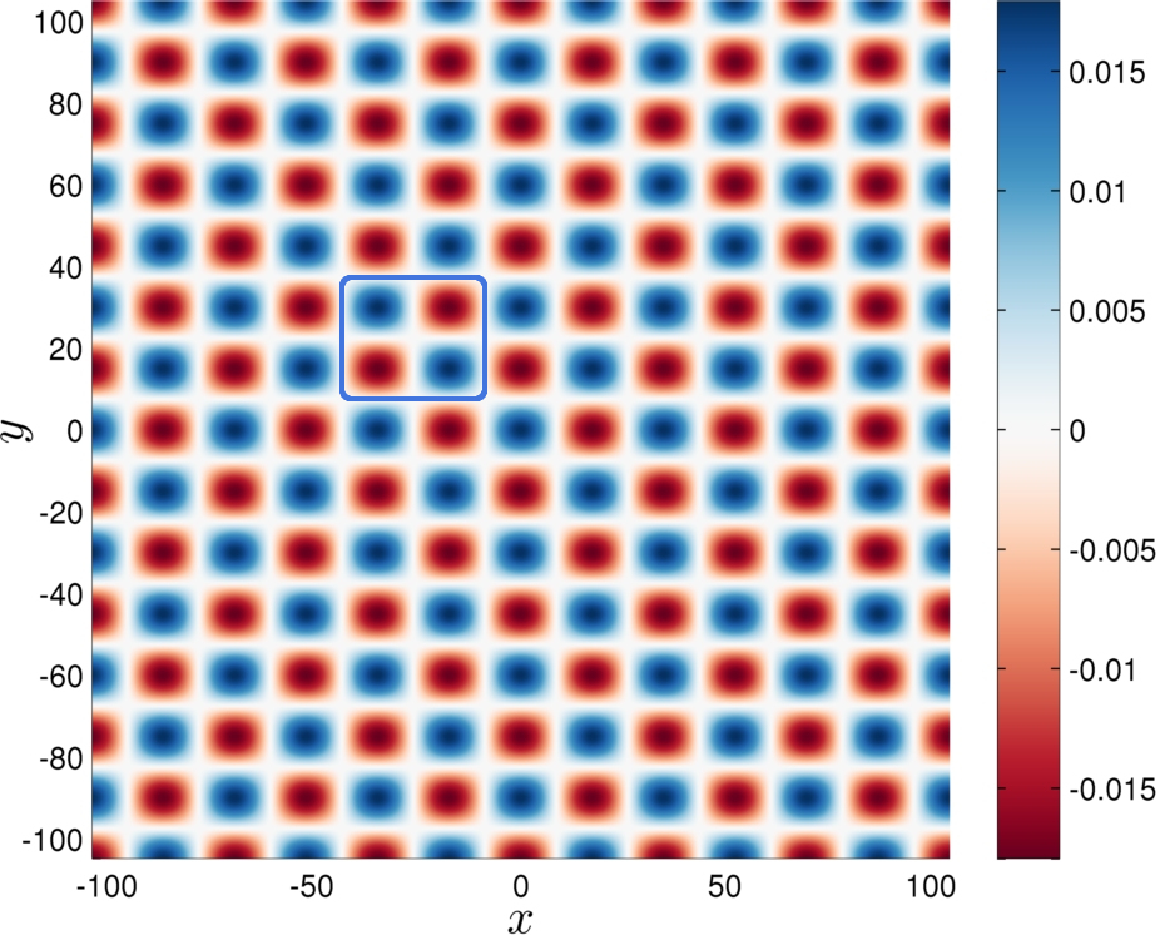}}
  \subfigure[]{%
  \includegraphics[width=0.63\textwidth]{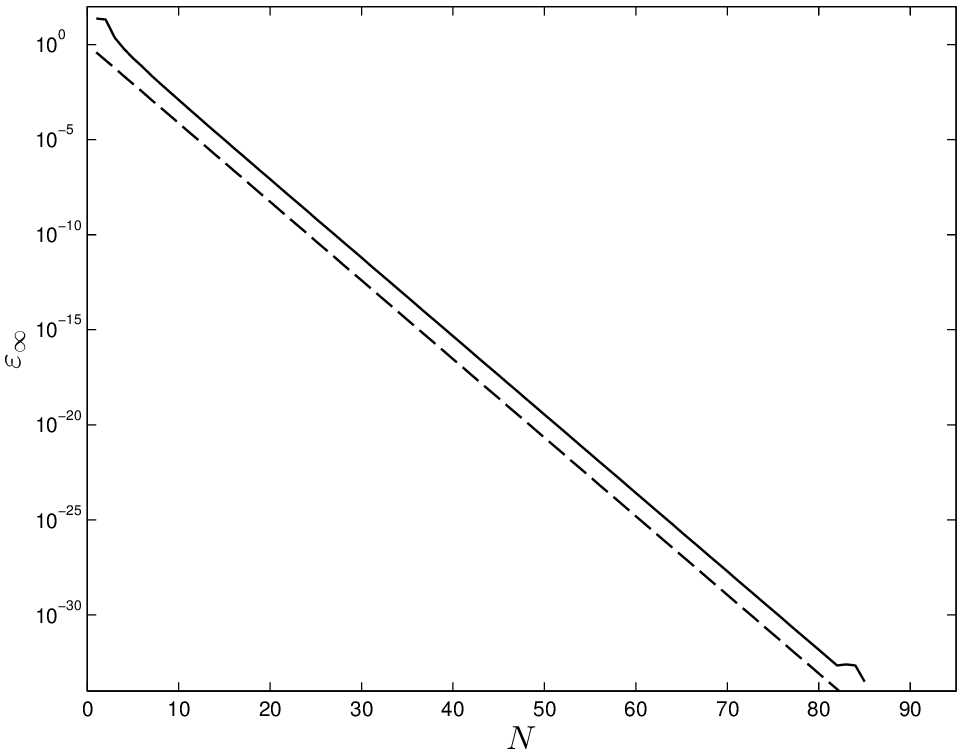}}
  \caption{\small\em (a) A bi-periodic wave pattern $B\,(x,\,y)\,$, $\Omega\ =\ [\,-105,\, 105]^{\,2}$ and $\omega\ =\ 0.012\,$, $1\,024 \times 1\,024$ \textsc{Fourier} modes. The box delimits the elementary $3\times 3$ discrete pattern identified to describe $B$ by substitutions on the two-letter alphabet $\{r,\, b\}$ used to code red spots (negative values far from zero), blue spots (positive values far from zero)respectively (b) Convergence test of the \textsc{Petviashvili} scheme in multi-precision arithmetics: (Solid line) $L_{\,\infty}$ norm between two successive iterations and (dash line) exponential fit $\ue^{\,-0.095 N}\,$.}
  \label{fig:f0}
\end{figure}

For example, consider the domain $\Omega\ \eqdef\ \Bigl[\,-\dfrac{\ell_{\,x}}{2},\, \dfrac{\ell_{\,x}}{2}\,\Bigr] \times \Bigl[\,-\dfrac{\ell_{\,y}}{2},\, \dfrac{\ell_{\,y}}{2}\,\Bigr]\ \subset\ \R^{\,2}$ to be a square with the side length equal to $210$ ($\Omega\ \equiv\ [\,-105,\, 105\,]^{\,2}$). For $\omega\ =\ 0.012\,$, the \textsc{Petviashvili} method on a $1\,024 \times 1\,024$ \textsc{Fourier} grid yields the strictly periodic regular pattern shown in Figure~\ref{fig:f0}(\textit{a}). The convergence is attained in $N\sim 80$ iterations as clearly seen in Figure~\ref{fig:f0}(\textit{b}) and the associated error $L_{\,\infty}\ \sim\ \O(10^{\,-33})\,$, and the residual $E_{\,r}\ \sim\ \O(10^{\,-32})\,$. Note that $L_{\,\infty}\ \sim\ \ue^{\,-0.095\, N}$ decays exponentially in agreement with the theoretical geometric convergence rate \cite{Alvarez2014}. For the sake of efficiency, the numerical solutions presented below are computed on the same \textsc{Fourier} grid using standard double-precision arithmetics. It is verified that residual errors are within the prescribed tolerance parameter $\varepsilon\ \sim\ 10^{\,-15}$ and never exceed $10\,\varepsilon\,$.

\begin{figure}
  \centering
  \includegraphics[width=0.75\textwidth]{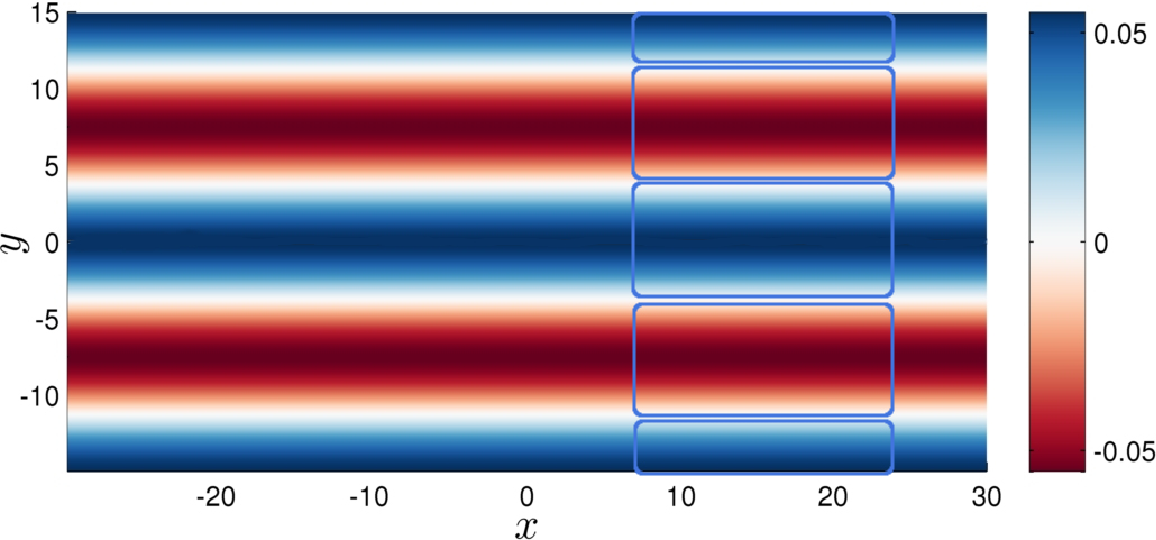}
  \caption{\small\em A bi-periodic wave pattern $B\,(x,\,y)\,$, $\Omega\ =\ [\,-30,\, 30\,] \times [\,-15,\, 15\,]$ and $\omega\ =\ 0.18\,$. The function $B$ can be described by the  $1\times 4$ discrete pattern $E\ =\ {}^t(r,\, b,\, r,\, b)$ on the two-letter alphabet $\{r,\, b\}$ (red spots code negative values far from zero and blue spots refer to positive values far from zero).}
  \label{fig:f1}
\end{figure}

\begin{figure}
  \centering
  \includegraphics[width=0.75\textwidth]{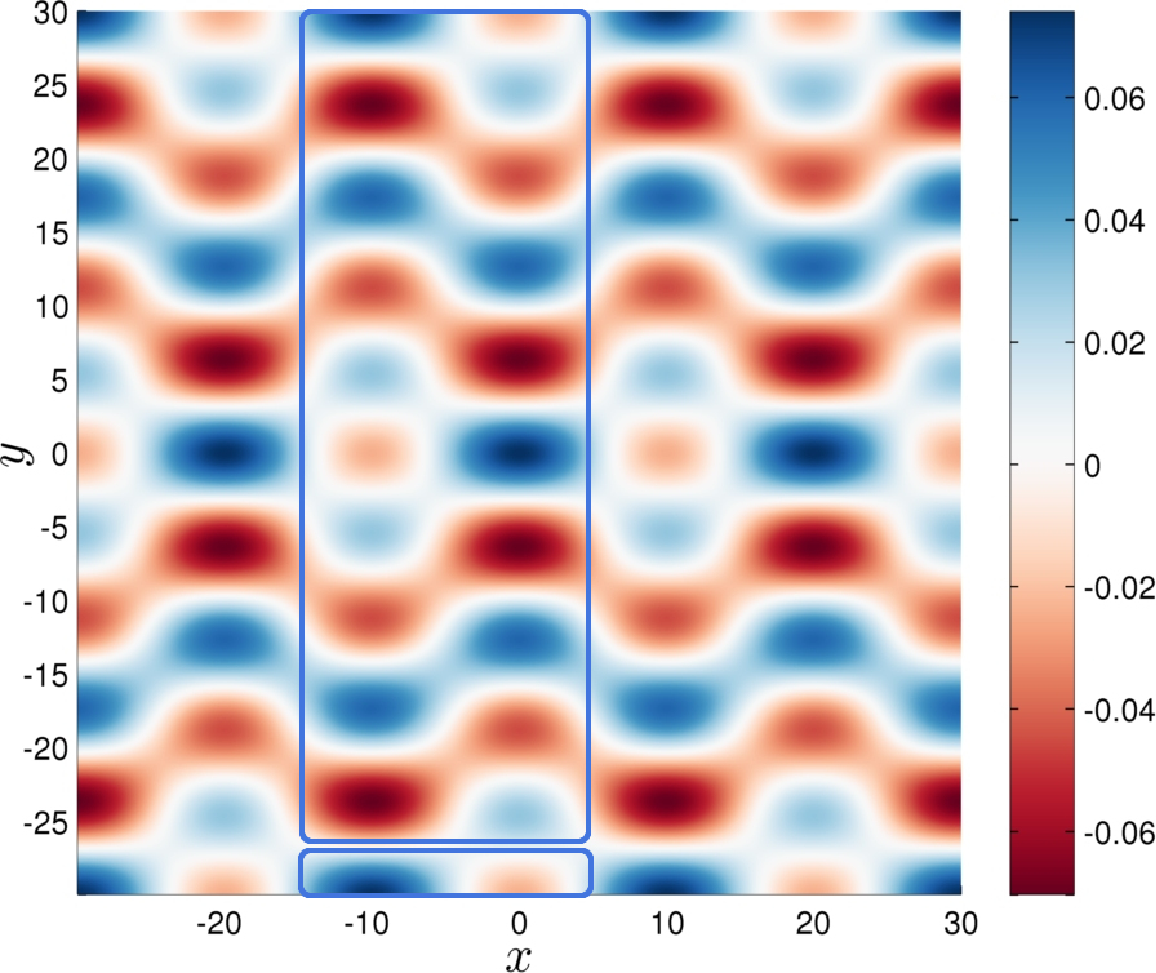}
  \caption{\small\em A bi-periodic wave pattern $B\,(x,\,y)\,$, $\Omega\ =\ [-30,\, 30]^{\,2}$ and $\omega\ =\ 0.18\,$. The box delimits the elementary $6\times 10$ discrete pattern of blue ($b$), red ($r$) and white ($w$, denoting the values around zero) spots identified to describe $B$ by substitutions on the three-letter alphabet $\{r, b, w\}$. For symbolic coding see caption of Figure~\ref{fig:f0}.}
  \label{fig:f2}
\end{figure}

\begin{figure}
  \centering
  \begin{overpic}[width=0.99\textwidth]{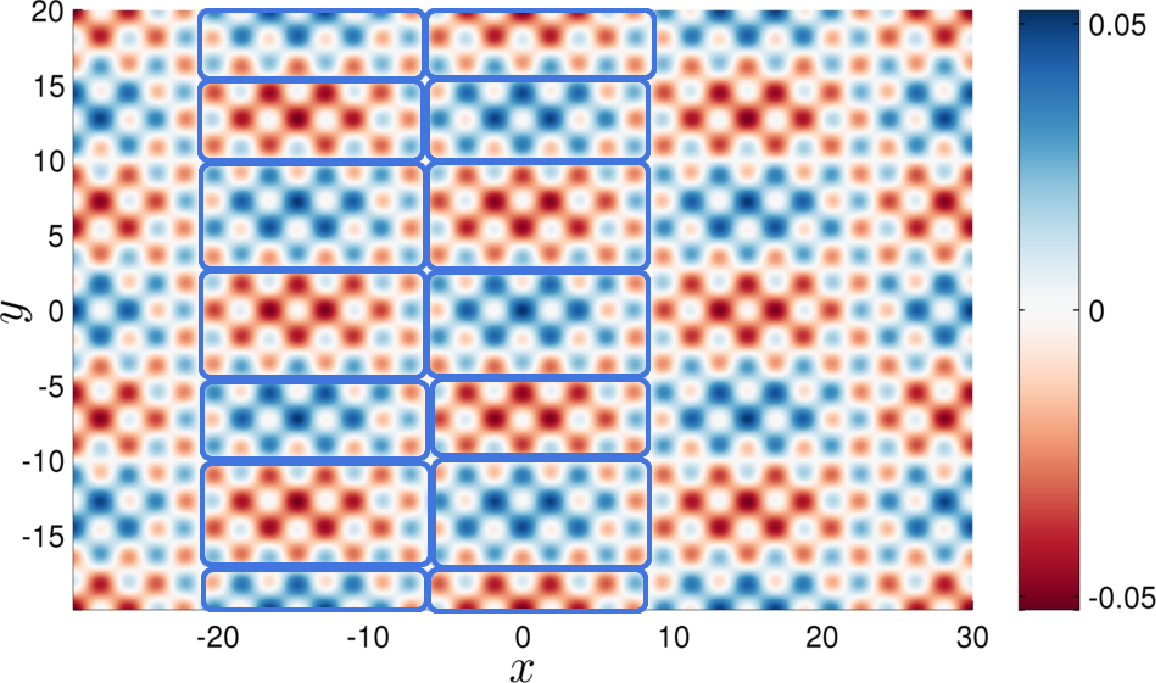}
  \put (24,54.5) {\Large$\sigma(b)$}
  \put (43,54.5) {\Large$\sigma(a)$}
  \put (24,47.5) {\Large$\sigma(c)$}
  \put (43,47.5) {\Large$\sigma(d)$}
  \put (24,39.5) {\Large$\sigma(\bar{b})$}
  \put (43,39.5) {\Large$\sigma(\bar{a})$}
  \put (24,30)   {\Large$\sigma(\bar{a})$}
  \put (43,30)   {\Large$\sigma(\bar{b})$}
  \put (24,22)   {\Large$\sigma(d)$}
  \put (43,22)   {\Large$\sigma(c)$}
  \put (24,14)   {\Large$\sigma(a)$}
  \put (43,14)   {\Large$\sigma(b)$}
  \end{overpic}
  \caption{\small\em A bi-periodic wave pattern $B\,(x,\,y)\,$, $\Omega\ =\ [\,-30,\, 30\,]\times[\,-20,\, 20\,]$ and $\omega = 0.18$. The box delimits the elementary $32\times 22$discrete pattern of blue ($b$) red ($r$) and white ($w$) spots identified to describe $B$ by substitutions on the three-letter alphabet $\{r,\, b,\, w\}\,$. For symbolic coding see caption of Figure~\ref{fig:f1}.}
  \label{fig:f3}
\end{figure}

\begin{figure}
  \centering
  \includegraphics[width=0.99\textwidth]{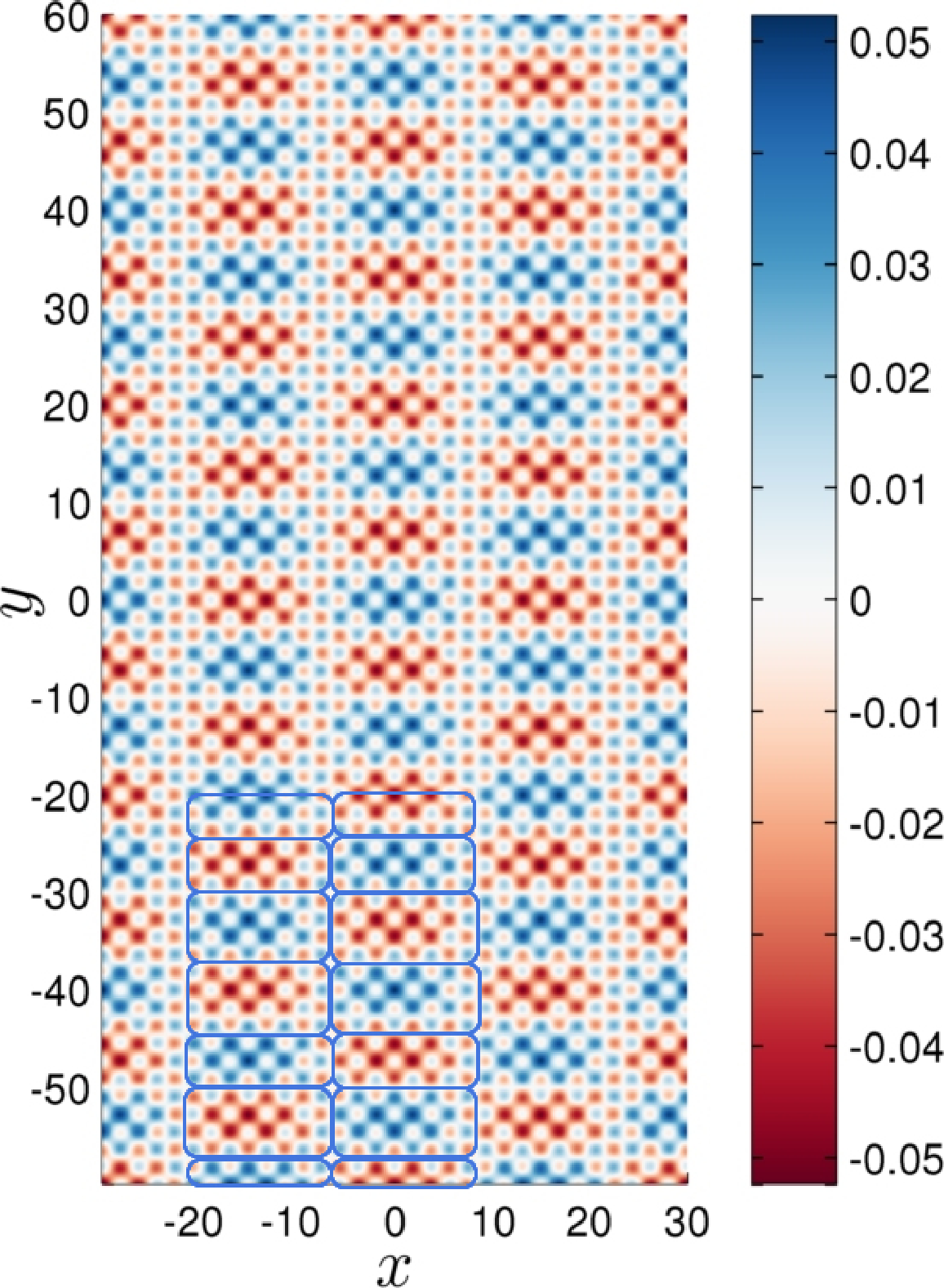}
  \caption{\small\em A bi-periodic wave pattern $B\,(x,\,y)\,$, $\Omega\ =\ [\,-30,\, 30\,]\times[\,-60,\, 60\,]$ and $\omega\ =\ 0.18\,$. The box delimits the elementary $32\times 66$ discrete pattern of blue ($b$) red ($r$) and white ($w$) spots identified to describe $B$ by substitutions on the three-letter alphabet $\{r,\, b,\, w\}\,$. For symbolic coding see Figure~\ref{fig:f0}.}
  \label{fig:f4}
\end{figure}

\begin{figure}
  \centering
  \includegraphics[width=0.99\textwidth]{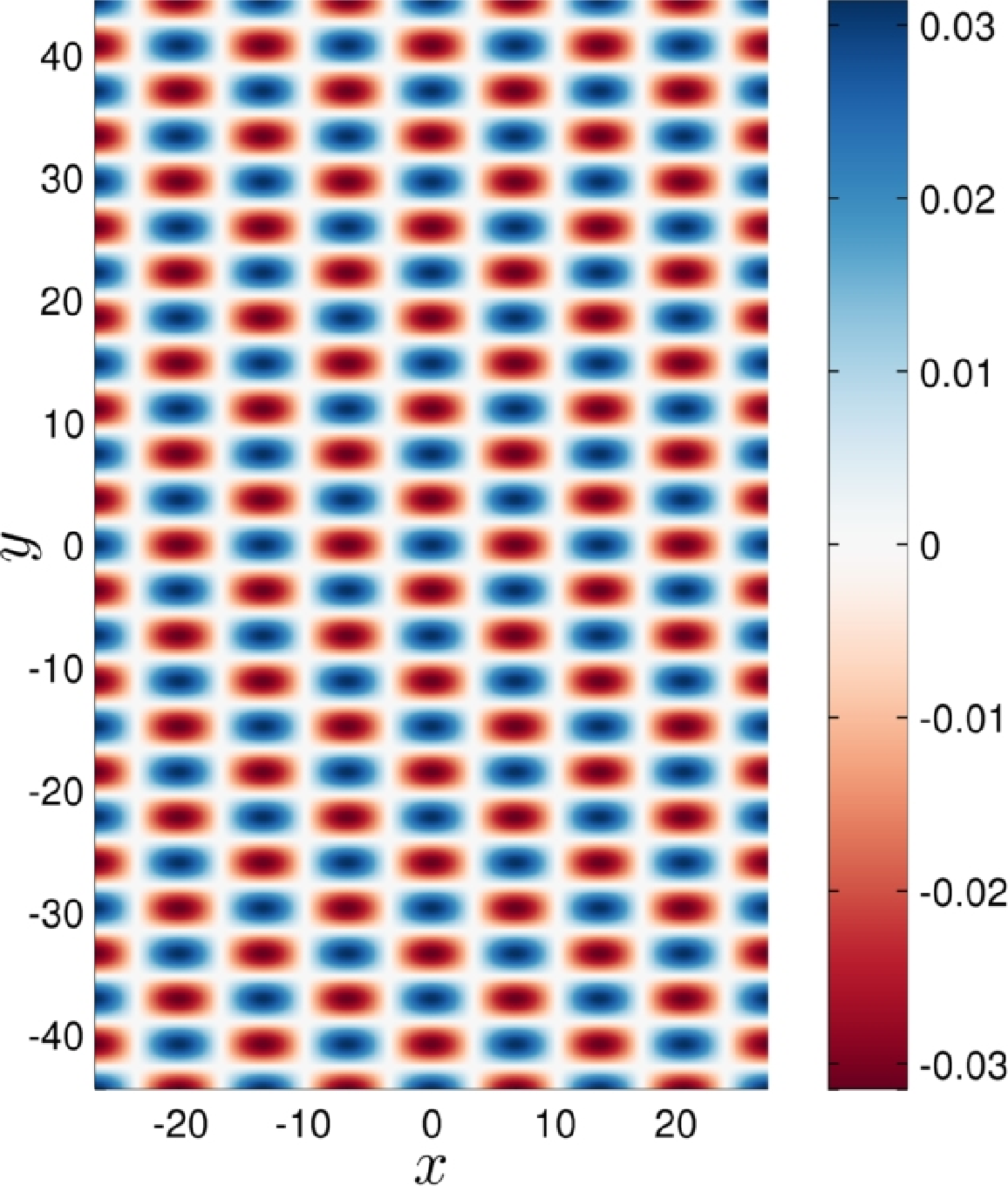}
  \caption{\small\em A bi-periodic wave pattern $B\,(x,\,y)\,$, $\Omega\ =\ \bigl[\,-\frac{55}{2},\, \frac{55}{2}\,\bigr]\times\bigl[\,-\frac{89}{2},\, \frac{89}{2}\,\bigr]$ and $\omega\ =\ 0.51\,$.}
  \label{fig:f5}
\end{figure}

\begin{figure}
  \centering
  \includegraphics[width=0.99\textwidth]{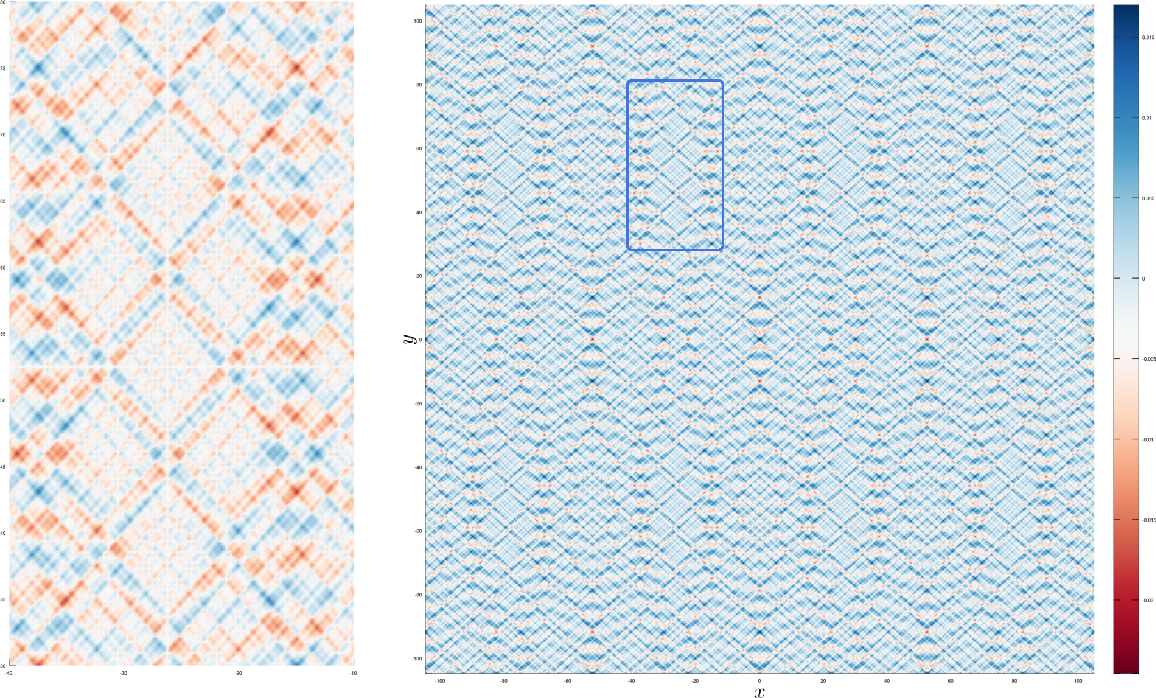}
  \caption{\small\em (Right) a bi-periodic wave pattern $B\,(x,\,y)\,$, $\Omega\ =\ [\,-105,\, 105]^{\,2}$ and $\omega\ =\ 1.3295\,$; (Left) zoom on the sub-region $[\,-40,\, -10\,] \times [\,30,\, 80\,]\,$.}
  \label{fig:f6}
\end{figure}


\section{Symbolic coding}
\label{sec:descr}

Symbolic dynamics was used by \textsc{Sturm} \cite{Sturm1836} and \textsc{Morse} \& \textsc{Hedlund} \cite{Morse1940} in order to study trajectories of special classes of ODEs and PDEs. This powerful method was built for $1-$D trajectories only. In this Section we extend the symbolic coding technique to describe $2-$D standing waves arising in the \acs{hNLS} equation. For this purpose the theory of bi-dimensional words and tilings is exploited. The key tool in tiling theory is the substitution operation that consists in replacing letters by certain groups of letters and constructing iteratively words on a finite alphabet.

The theory of bi-dimensional words and tilings (see \cite{Berthe2000, DeBruijn1981}) can be exploited to describe standing wave patterns of the \acs{hNLS} equation. The key tool in tiling theory is the substitution operation that consists in replacing letters by group of letters and constructing iteratively words on a finite alphabet. For example, the \textsc{Fibonacci} substitution $\sigma\,(a)\ =\ a\,b$ and $\sigma\,(b)\ =\ a$ constructs a fixed point $F$ (limit of infinite substitutions) with a non-periodic structure (\cite{Arnoux1991, Lind1995, Allouche2003} such that $\sigma\,(F)\ =\ F\,$. This \textsc{Fibonacci} word constitutes a model for $1-$D quasi-crystals in physics (see \cite{Senechal1995}) and a coding of a discrete line with irrational slope in discrete geometry (see \cite{Vuillon2003}). The \textsc{Fibonacci} word is a classical example of \textsc{Sturmian} words (\cf \cite{Vuillon2003}). The first link between PDEs and symbolic dynamics appeared in the work of \textsc{Sturm} (1836) and his famous zero separation theorem \cite{Sturm1836}. This theorem could be reformulated in terms of the symbolic dynamics. Indeed, consider a second order differential equation:
\begin{equation*}
  \od{^{\,2}\,y}{x^{\,2}}\ +\ \phi\,(x)\,y\ =\ 0\,,
\end{equation*}
where $\phi\,(x)$ is a continuous periodic function with period $1\,$. If $k_{\,n}\ \in\ \mathds{N}_{\,0}$ is the number of zeros in the interval $[\,n,\, n\ +\ 1)\,$. Then, the infinite word
\begin{equation*}
  1\,0^{\,k_{\,0}}\,1\,0^{\,k_{\,1}}\,1\,\ldots
\end{equation*}
is a \textsc{Sturmian} word or an ultimately periodic word. Hence, the connection between PDEs and symbolic dynamics is old and deep. That is why it seems natural to us to apply it to the description of bi-periodic standing wave patterns in the \acs{hNLS} equation.

The method of substitutions empowers the construction of either non-periodic or periodic words $G$ as fixed points of the mapping $G\ =\ \sigma^{\,n}\,(a)$ as $n\ \to\ +\infty\,$, where $a$ is a letter (see \cite{Lind1995}). On the one hand, non-periodic infinite words can be constructed iteratively by the use of morphism properties. In particular, a substitution $\sigma$ applied to a word $w\ =\ w_{\,1}\,w_{\,2}\cdots\, w_{\,n}$ is in fact the image of the substitution applied to each letter of $w\,$, that is $\sigma\,(w_{\,1}\,w_{\,2}\,\cdots\,w_{\,n})\ =\ \sigma\,(w_{\,1})\,\sigma\,(w_{\,2})\,\cdots\,\sigma\,(w_{\,n})\,$, where $w_{\,i}$ are letters of the alphabet. As an example, the first few iterations on the letter $a$ using the \textsc{Fibonacci} substitution are given by $\sigma^{\,1}\,(a)\ =\ \sigma\,(a)\ =\ a\,b\,$, $\sigma^{\,2}\,(a)\ =\ \sigma\,(\sigma\,(a))\ =\ \sigma\,(a\,b)\ =\ \sigma\,(a)\, \sigma\,(b)\ =\ a\,b\,a\,$, $\sigma^{\,3}\,(a)\ =\ \sigma\,(\sigma^{\,2}\,(a))\ =\ \sigma\,(a\,b\,a)\ =\ \sigma\,(a)\,\sigma\,(b)\,\sigma(a)\ =\ a\,b\,a\,a\,b\,$, and so on (see \cite{Arnoux1991, Vuillon2003}). Note that each word $\sigma^{\,n}\,(a)$ is the beginning of the next word $\sigma^{\,n+1}\,(a)\,$. Thus, as $n\ \to\ +\infty\,$, a non-periodic fixed point is constructed, \viz 
\begin{multline*}
  \sigma\,(F)\ =\ F\ =\ a\,b\,a\,a\,b\,a\,b\,a\,a\,b\,a\,a\,b\,a\,b\,a\,a\,b\,a\ldots\\
  b\,a\,a\,b\,a\,a\,b\,a\,b\,a\,a\,b\,a\,a\,b\,a\,b\,a\,a\,b\,a\,b\,a\,a\,b\,a\,a\,b\,a\,b\,a\,a\,b\,a\,b\,a\,a\,b\,a\,a\,b\,a\,b\,a\,a\,b\,a\,\cdots\,.
\end{multline*}

On the other hand, a periodic word can be constructed by means of the substitution $\sigma\,(a)\ =\ a\,b$ and $\sigma\,(b)\ =\ a\,b$ by repetition of the pattern $a\,b\,$. Indeed $\sigma^1\,(a)\ =\ \sigma\,(a)\ =\ a\,b\,$, $\sigma^{\,2}\,(a)\ =\ \sigma\,(\sigma\,(a))\ =\ \sigma\,(a\,b)\ =\ \sigma\,(a)\,\sigma\ (b)\ =\ a\,b\,a\,b\,$, $\sigma^{\,3}\,(a)\ =\ \sigma\,(\sigma^{\,2}\,(a))\ =\ \sigma\,(a\,b\,a\,b)\ =\ \sigma\,(a)\,\sigma\,(b)\,\sigma\,(a)\,\sigma\,(b)\ =\ a\,b\,a\,b\,a\,b\,a\,b\,$, and so on. The fixed point is periodic and given by 
\begin{multline*}
  \sigma\,(P)\ =\ P\ =\ a\,b\,a\,b\,a\,b\,a\,b\,a\,b\,a\,b\,a\,b\,a\,b\,a\,b\,a\ldots\\
  b\,a\,b\,a\,b\,a\,b\,a\,b\,a\,b\,a\,b\,a\,b\,a\,b\,a\,b\,a\,b\,a\,b\,a\,b\,a\,b\,a\,b\,a\,b\,a\,b\,a\,b\,a\,\cdots\,.
\end{multline*}

In order to code periodic spatial patterns of the \acs{hNLS} equation, one can exploit bi-dimensional words (see \cite{Frougny2005, Berthe2000, Lind1995}) and proper substitutions (see \cite{Arnoux1991, Berthe2000, Frougny2005}). For example, the bi-dimensional \textsc{Thue}--\textsc{Morse} substitution (see \cite{Allouche2003}) is defined by
\begin{equation*}
  \sigma\,(a)\ =\ 
\begin{pmatrix}
  a & b \\ 
  b & a
\end{pmatrix}
\qquad \mbox{and} \qquad 
\sigma\,(b)\ =\ 
\begin{pmatrix}
  b & a \\ 
  a & b 
\end{pmatrix}\,.
\end{equation*}

One can just iterate a finite number of substitutions, \viz $\sigma^{\,j}\,(a)$ with $j$ finite, and obtain a finite world. For example, the first two iterations of the \textsc{Thue}--\textsc{Morse} substitution yield, respectively, the two words
\begin{equation*}
\sigma^{\,1}\,(a)\ =\ \sigma\,(a)\ =\ 
\begin{pmatrix}
  a & b \\ 
  b & a
\end{pmatrix}
\end{equation*}
and 
\begin{equation*}
\sigma^{\,2}\,(a)\ =\ \sigma\,
\begin{pmatrix}
  a & b \\ 
  b & a
\end{pmatrix}\ =\ 
\begin{pmatrix}
  \sigma\,(a) & \sigma\,(b) \\ 
  \sigma\,(b) & \sigma\,(a)
\end{pmatrix}\ =\ 
\begin{pmatrix}
  a & b & b & a \\
  b & a & a & b \\
  b & a & a & b \\
  a & b & b & a
\end{pmatrix}\,.
\end{equation*}

In the limit of infinite iterations, an infinite word is obtained in the form of the non-periodic fixed point
\begin{equation*}
  \sigma\,(M)\ =\ M\ =\ 
\begin{pmatrix}
  a & b & b & a & b & a\; \cdots \\
  b & a & a & b & a & b\; \cdots \\
  b & a & a & b & a & b\; \cdots \\
  a & b & b & a & b & a\; \cdots \\
  b & a & a & b & a & b\; \cdots \\
  a & b & b & a & b & a\; \cdots \\
\vdots & \vdots & \vdots & \vdots & \vdots & \vdots ~ \ddots
\end{pmatrix}\,.
\end{equation*}

Hereafter, a discrete description of several spatial patterns $B\,(x,\,y)$ of the \acs{hNLS} equation obtained via the \textsc{Petviashvili} method \cite{Petviashvili1976} is presented. Each continuous pattern is coded on a finite alphabet using proper substitutions iterated on a discrete elementary pattern of letters.

Consider the bi-periodic pattern $B$ of the \acs{hNLS} equation shown in Figure~\ref{fig:f0}. This can be easily described by repetition of the elementary $3 \times 3$ discrete pattern delimited by a box in the same Figure. To do so, coding with a three--letter alphabet is used since the pattern is characterized by a discrete structure with three different elementary spots. The letters $r\,$, $b$ and $w$ are used to code red spots (negative values far from zero), blue spots (positive values far from zero) and white spots (values around zero) respectively.

As clearly seen in Figure~\ref{fig:f0}, the three spots are arranged on a rectangular sub-region of the $\GZ^{\,2}$ grid. Thus, the continuous periodic pattern $B$ can be easily described by bi-dimensional words constructed using the two simple substitutions
\begin{equation*}  
  \sigma\,(r)\ =\ 
\begin{pmatrix}
  r & b \\ 
  b & r
\end{pmatrix} \qquad \mbox{and} \qquad 
\sigma\,(b)\ =\ 
\begin{pmatrix}
 r & b \\ 
 b & r 
\end{pmatrix}\,.
\end{equation*}

By iterating the above substitutions yields the strictly periodic word
\begin{equation*}
  \sigma\,(R)\ =\ R\ =\ 
\begin{pmatrix}
  r & b & r & b\; \cdots\\
  b & r & b & r\; \cdots\\
  r & b & r & b\; \cdots \\
  b & r & b & r\; \cdots\\
  \vdots & \vdots & \vdots & \vdots ~ \ddots
\end{pmatrix}\,,
\end{equation*}
which describes the continuous wave pattern of Figure~\ref{fig:f0}.

Consider now the wave pattern solutions for $B\,(x,\,y)$ obtained for $\omega\ =\ 0.18$ on various rectangular grid sizes. For $\Omega\ =\ [\,-30,\, 30\,]\times [\,-15,\, 15\,]\,$, $B$ can be described by a $1 \times 4$ discrete pattern delimited by a box in Figure~\ref{fig:f1} and coded as $E\ =\ {}^{\top}(r,\,b,\,r,\,b)\,$, where the superscript ${}^{\top}$ denotes matrix transposition. The code $E$ can be decomposed as $\sigma\,(e)\ =\ {}^{\top}\,(r,\, b)$ and the initial pattern $E$ is given by the substitution $\sigma$ applied to $E\ =\ {}^{\top}(e,\, e)$ that is
\begin{equation*}
  \sigma\,(E)\ =\ {}^{\top}\bigl(\sigma\,(e),\; \sigma\,(e)\bigr)\ =\ {}^{\top}\bigl(r,\, b,\, r,\, b\bigr)\,.
\end{equation*}

For $\Omega\ =\ [\,-30,\, 30\,]\times [\,-30,\, 30\,]\,$, the associated continuous pattern $B$ is shown in Figure~\ref{fig:f2}. A box delimits the elementary  $6 \times 10$  discrete pattern $F$ that can describe $B$ by successive substitutions, \viz  
\begin{equation*}
  F\ =\ \begin{pmatrix}
  b & w & b & w & b & w \\
  r & w & r & w & r & w \\
  b & r & b & r & b & r \\
  r & b & r & b & r & b \\
  w & r & w & r & w & r \\
  w & b & w & b & w & b \\
  w & r & w & r & w & r \\
  r & b & r & b & r & b \\
  b & r & b & r & b & r \\
  r & w & r & w & r & w
\end{pmatrix}\,.
\end{equation*}

This discrete pattern can be decomposed as
\begin{equation*}
  \sigma\,(f)\ =\  
  {}^{\top}\begin{pmatrix}
    b & r & b & r & w & w & w & r & b & r \\
    w & w & r & b & r & b & r & b & r & w
\end{pmatrix}\,.
\end{equation*}
As a result, the initial pattern $F$ is given by the substitution $\sigma$ applied to $F\ =\ (f~f~f)\,$, that is
\begin{equation*}
  \sigma\,(F)\ =\ \bigl(\sigma\,(f) ~ \sigma\,(f) ~ \sigma\,(f)\bigr)\ =\
\begin{pmatrix}
  b & w & b & w & b & w \\
  r & w & r & w & r & w \\
  b & r & b & r & b & r \\
  r & b & r & b & r & b \\
  w & r & w & r & w & r \\
  w & b & w & b & w & b \\
  w & r & w & r & w & r \\
  r & b & r & b & r & b \\
  b & r & b & r & b & r \\
  r & w & r & w & r & w
\end{pmatrix}\,.
\end{equation*}

For the larger domain $\Omega\ =\ [\,-30,\, 30\,]\times [\,-20,\, 20\,]\,$, the associated standing wave pattern $B\,(x,\,y)$ is reported in Figure~\ref{fig:f3}. An elementary $32\times 22$ discrete pattern of blue ($b$), red ($r$) and white ($w$) spots is identified and delimited by a box to describe $B\,$. This can be obtained by four iterations of the following $6$ substitutions on the three-letter alphabet $\{r,\, b,\, w\}\,$, that is
\begin{equation*}
  \sigma\,(a)\ =\ 
  \begin{pmatrix}
    r & w & r & w & r & w & r & w \\
    w & r & w & r & w & r & w & w \\
    r & w & r & w & r & w & r & w \\
    w & w & w & w & w & w & w & w
  \end{pmatrix}\,,
\end{equation*}
\begin{equation*}
  \sigma\,(\bar{a})\ =\ 
  \begin{pmatrix}
    w & r & w & r & w & r & w & w \\
    r & w & r & w & r & w & r & w \\
    w & r & w & r & w & r & w & w \\
    w & w & w & w & w & w & w & w
  \end{pmatrix}\,,
\end{equation*}
\begin{equation*}
  \sigma\,(b)\ =\ 
  \begin{pmatrix}
    w & b & w & b & w & b & w & w \\
    b & w & b & w & b & w & b & w \\
    w & b & w & b & w & b & w & w \\
    w & w & w & w & w & w & w & w
  \end{pmatrix}\,,
\end{equation*}
\begin{equation*}
  \sigma\,(\bar{b})\ =\
  \begin{pmatrix}
    b & w & b & w & b & w & b & w \\
    w & b & w & b & w & b & w & w \\
    b & w & b & w & b & w & b & w \\
    w & w & w & w & w & w & w & w
  \end{pmatrix}\,,
\end{equation*}
\begin{equation*}
  \sigma\,(c)\ =\ 
  \begin{pmatrix}
    r & w & r & w & r & w & r & w \\
    w & r & w & r & w & r & w & w \\
    r & w & r & w & r & w & r & w
  \end{pmatrix}\,,
\end{equation*}
\begin{equation*}
  \sigma\,(d)\ =\ 
  \begin{pmatrix}
    w & b & w & b & w & b & w & w \\
    b & w & b & w & b & w & b & w \\
    w & b & w & b & w & b & w & w
  \end{pmatrix}\,.
\end{equation*}
As described above, this substitution is applied four times under toric constraints starting from the finite $4 \times 6$ pattern $G$ given by
\begin{equation*}
  G\ =\ 
  \begin{pmatrix}
    b & a & b & a \\
    c & d & c & d \\
    \bar{b} & \bar{a} & \bar{b} & \bar{a} \\
    \bar{a} & \bar{b} & \bar{a} & \bar{b} \\
    d & c & d & c \\
    a & b & a & b
  \end{pmatrix}\,.
\end{equation*}
Here, the difference between $\sigma\,(a)$ and $\sigma\,(\bar{a})$ resides in the interchange of the letters $r$ and $w$ in the first $3$ rows of the pattern except the last column, and similarly for $\sigma\,(b)$ and $\sigma\,(\bar{b})\,$. This switching between components is exactly the typical dynamical property that we expect for symbolic coding of patterns. For example, the first iteration $\sigma\,(G)$ yields
\begin{equation*}
  \sigma\,(G) = \sigma\,\left(\begin{array}{llllllllllllllll}
  b & a & b & a \\
  c & d & c & d \\
  \bar{b} & \bar{a} & \bar{b} & \bar{a} \\
  \bar{a} & \bar{b} & \bar{a} & \bar{b} \\
  d & c & d & c \\
  a & b & a & b
\end{array}\right)\ =\ 
  \begin{pmatrix}
  \sigma\,(b) & \sigma\,(a) & \sigma\,(b) & \sigma\,(a) \\
  \sigma\,(c) & \sigma\,(d) & \sigma\,(c) & \sigma\,(d) \\
  \sigma\,(\bar{b})         & \sigma\,(\bar{a}) & \sigma\,(\bar{b}) & \sigma\,(\bar{a}) \\
  \sigma\,(\bar{a})         & \sigma\,(\bar{b}) & \sigma\,(\bar{a}) & \sigma\,(\bar{b}) \\
  \sigma\,(d) & \sigma\,(c) & \sigma\,(d) & \sigma\,(c) \\
  \sigma\,(a) & \sigma\,(b) & \sigma\,(a) & \sigma\,(b)
  \end{pmatrix}\,.
\end{equation*}
The whole discrete description of the continuous pattern $B\,(x,\,y)$ follows after four iterations as
\begin{equation*}
  \sigma\,(G)\ =\
  \left(\begin{array}{lllllllllllllllllllllllllllllllllll}
  w & b & w & b & w & b & w & w & r & w & r & w & r & w & r & w\; \cdots \\
  b & w & b & w & b & w & b & w & w & r & w & r & w & r & w & w\; \cdots \\
  w & b & w & b & w & b & w & w & r & w & r & w & r & w & r & w\; \cdots \\
  w & w & w & w & w & w & w & w & w & w & w & w & w & w & w & w\; \cdots \\
  r & w & r & w & r & w & r & w & w & b & w & b & w & b & w & w\; \cdots \\
  w & r & w & r & w & r & w & w & b & w & b & w & b & w & b & w\; \cdots \\
  r & w & r & w & r & w & r & w & w & b & w & b & w & b & w & w\; \cdots \\
  b & w & b & w & b & w & b & w & w & r & w & r & w & r & w & w\; \cdots \\
  w & b & w & b & w & b & w & w & r & w & r & w & r & w & r & w\; \cdots \\
  b & w & b & w & b & w & b & w & w & r & w & r & w & r & w & w\; \cdots \\
  w & w & w & w & w & w & w & w & w & w & w & w & w & w & w & w\; \cdots \\
  w & r & w & r & w & r & w & w & b & w & b & w & b & w & b & w\; \cdots \\
  r & w & r & w & r & w & r & w & w & b & w & b & w & b & w & w\; \cdots \\
  w & r & w & r & w & r & w & w & b & w & b & w & b & w & b & w\; \cdots \\
  w & w & w & w & w & w & w & w & w & w & w & w & w & w & w & w\; \cdots \\
  \vdots & \vdots & \vdots & \vdots & \vdots & \vdots & \vdots & \vdots & \vdots & \vdots & \vdots & \vdots & \vdots & \vdots & \vdots & \vdots ~ \ddots
\end{array}\right)\,.
\end{equation*}

For $\Omega\ =\ [\,-30,\, 30\,]\times [\,-60,\, 60\,]\,$, the associated water elevation pattern $B$ can be described by the elementary $32 \times 66$ discrete pattern $K$ delimited by a box and shown in Figure~\ref{fig:f4}. In this case, the symbolic description of $K$ follows from the substitution $K\ =\ \sigma\,(H)\,$, where $\sigma$ is defined as
\begin{equation*}
  \sigma\,(a)\ =\ 
  \begin{pmatrix}
    r & w & r & w & r & w & r & w \\
    w & r & w & r & w & r & w & w \\
    r & w & r & w & r & w & r & w \\
    w & w & w & w & w & w & w & w
  \end{pmatrix}\,,
\end{equation*}
\begin{equation*}
  \sigma\,(\bar{a})\ =\
  \begin{pmatrix}
    w & r & w & r & w & r & w & w \\
    r & w & r & w & r & w & r & w \\
    w & r & w & r & w & r & w & w \\
    w & w & w & w & w & w & w & w
  \end{pmatrix}\,,
\end{equation*}
\begin{equation*}
  \sigma\,(b)\ =\ 
  \begin{pmatrix}
    w & b & w & b & w & b & w & w \\
    b & w & b & w & b & w & b & w \\
    w & b & w & b & w & b & w & w \\
    w & w & w & w & w & w & w & w
  \end{pmatrix}\,,
\end{equation*}
\begin{equation*}
  \sigma\,(\bar{b})\ =\ 
  \begin{pmatrix}
    b & w & b & w & b & w & b & w \\
    w & b & w & b & w & b & w & w \\
    b & w & b & w & b & w & b & w \\
    w & w & w & w & w & w & w & w
  \end{pmatrix}\,,
\end{equation*}
\begin{equation*}
  \sigma\,(c)\ =\
  \begin{pmatrix}
    r & w & r & w & r & w & r & w \\
    w & r & w & r & w & r & w & w \\
    r & w & r & w & r & w & r & w
  \end{pmatrix}\,,
\end{equation*}
\begin{equation*}
  \sigma\,(d)\ =\ 
  \begin{pmatrix}
    w & b & w & b & w & b & w & w \\
    b & w & b & w & b & w & b & w \\
    w & b & w & b & w & b & w & w
  \end{pmatrix}\,,
\end{equation*}
and applied to the following $4 \times 12$ pattern $H$ under toric constraints, \viz
\begin{equation*}
  H\ =\ 
  \begin{pmatrix}
    G \\
    G
  \end{pmatrix}\ =\ {}^{\top}\left(
  \begin{array}{llllllllllll}
    b & c & \bar{b} & \bar{a} & d & a & b & c & \bar{b} & \bar{a} & d & a \\
    a & d & \bar{a} & \bar{b} & c & d & a & d & \bar{a} & \bar{b} & c & b \\
    b & c & \bar{b} & \bar{a} & d & a & b & c & \bar{b} & \bar{a} & d & a \\
    a & d & \bar{a} & \bar{b} & c & b & a & d & \bar{a} & \bar{b} & c & b
  \end{array}\right)\,.
\end{equation*}

Note that $H$ is made of two copies of the discrete structure $G$ identified for the pattern $B\,(x,\,y)$ relative to the domain $\Omega\ =\ [\,-30,\, 30\,]\times [\,-20,\, 20\,]\,$, see Figure~\ref{fig:f3}. Thus, $\sigma\,(H)$ is  exactly the substitution applied to two copies of $G\,$, \viz $\sigma\,(H)\ =\ {}^{\top}\bigl(\,\sigma\,(G),\, \sigma\,(G)\,\bigr)\,$. This suggests that for a given value of $\omega$ the patterns associated to various domains $\Omega$ have in common the same substitutions. The next goal will be to construct all the substitutions according to special values of $\omega$ and various domains $\Omega\,$. These techniques capture the structure of the discrete patterns and the nature of the dynamical system.

Future research aims at finding a sequence of increasing domain sizes whose associated discrete patterns share the same substitution. This leads to construct periodic orbits and fixed points of the dynamics. For example, Figure~\ref{fig:f5} shows a patterns that may be generated by substitutions of elementary cell patterns. Moreover, for given domain size, one could explore which frequencies $\omega$ gives the same kind of substitutions and try to explain this regularity by the arithmetic nature of $\omega\,$. More precisely, suppose that for a given $\omega$ one identifies two discrete patterns generated by repeated substitutions $\sigma^{\,k}\,(g)$ and $\sigma^{\,\ell}\,(g)\,$, respectively, with $k$ and $\ell$ as the number of repetitions. Then, one may find a sequence of increasing domain sizes $(\Omega_{\,i})_{\,i\, \in\, I}$ and try to associate substitutive patterns of the form $\sigma^{\,i}\,(g)\,$. This could yield characteristic scales with invariance of patterns and could suggest that the dynamics is given by a coding of iterated substitutions of the form $(\sigma^{\,i}\,(g)))_{\,i\,\in\,I}\,$. A fixed point $G$ such that $\sigma\,(G)\ =\ G$ could then exist (see \cite{Morse1940, Lind1995}). Furthermore, note that there is a link between the complexity of the patterns and the decomposition in prime factors of the domain $\Omega$ size. For example, the domain $\Omega\ =\ \Bigl[\,-\dfrac{\ell}{2},\, \dfrac{\ell}{2}\,\Bigr] \times \Bigl[\,-\dfrac{\ell}{2},\, \dfrac{\ell}{2}\,\Bigr]$ with length $\ell\ =\ 2 \times 3 \times 5 \times 7\ =\ 210$ is the product of the four first prime numbers and the associated complex pattern can be seen in Figure~\ref{fig:f6}. The apparent complexity of the solution may be explained by a combinations of well chosen substitutions. One expects that the larger domain $\Omega$ with length $\ell\ =\ 2 \times 3 \times 5 \times 7 \times 11\ =\ 2310$ leads to even more complicated pattern solutions. Indeed, this decomposition in prime factors increases the number of divisors of the pattern size and, thus, the complexity of patterns. To summarize, this part presents the first step for a symbolic dynamic coding of patterns arising from non linear waves. We will use in next works these substitutions in order to understand the complexity of such patterns. Furthermore, as the study uses discrete patterns note that there is a link between the complexity of the patterns and the decomposition in prime factors of the domain $\Omega$ size.


\section{Dynamics of perturbed standing waves}
\label{sec:descr}

\subsection{Pseudo-spectral scheme}

A highly accurate \textsc{Fourier}-type pseudo-spectral method \cite{Trefethen2000, Boyd2000} will be used to solve the unsteady \acs{hNLS} equation \eqref{eq:adim} in order to investigate the dynamics of a perturbed standing wave pattern. To do so, equation \eqref{eq:adim} is recast in the following form:
\begin{equation}\label{eq:oper}
  A_{\,t}\ +\ \ui\,\L\cdot A\ =\ \N\,(A)\,,
\end{equation}
where the operators $\L$ and $\N$ are defined such as:
\begin{equation*}
  \L\ \eqdef\ \partial_{\,x\,x}\ -\ \partial_{\,y\,y}\,, \qquad \N\,:\ A\ \mapsto\ -2\,\ui\,\abs{A}^{\,2}\,A\,.
\end{equation*}
With this setting, \eqref{eq:oper} is discretized by applying the $2-$D \textsc{Fourier} transform $\F$ in the spatial variables $(x,\, y)\,$. The nonlinear terms are computed in physical space, while spatial derivatives are computed spectrally in \textsc{Fourier} space. The standard $3/2$ rule is applied for anti-aliasing \cite{Trefethen2000, Clamond2001, Fructus2005}. The transformed variables will be denoted by $\hat{A}\,(t,\, \k)\ =\ \F\{A\,(t,\,x,\,y)\}\,$, with $\k\ =\ (k_{\,x},\, k_{\,y})$ being the \textsc{Fourier} transform parameter.

In order to improve the stability of the time discretization procedure, the linear part of the operator is integrated exactly\footnote{Sometimes this method in the literature is called the `integrating factor method' (\cf \cite{Trefethen2000}).} by a change of variables \cite{Lawson1967, Milewski1999, Fructus2005} that yields
\begin{equation*}
  \hat{\A}_{\,t}\ =\ \ue^{\,(t\ -\ t_{\,0})\,\L}\cdot\N\,\Bigl\{\,\ue^{\,-(t\ -\ t_{\,0})\,\L}\cdot\hat{\A}\,\Bigr\}\,, \qquad \hat{\A}\,(t) \eqdef\ \ue^{\,(t\ -\ t_{\,0})\,\L}\cdot\hat{A}\,(t)\,, \qquad \hat{\A}\,(t_{\,0})\ =\ \hat{A}\,(t_{\,0})\,.
\end{equation*}
The exponential matrix $\hat{\L}$ is explicitly computed in \textsc{Fourier} space as
\begin{equation*}
  \ue^{\,(t\ -\ t_{\,0})\,\hat{\L}}\ =\ \ue^{\,-\ui\,(\,k_{\,x}^{\,2}\ -\ k_{\,y}^{\,2})\,(t\ -\ t_{\,0})}\,.
\end{equation*}
Finally, the resulting system of ODEs is discretized in time by the \textsc{Verner}'s embedded adaptive $9(8)$ \textsc{Runge}--\textsc{Kutta} scheme \cite{Verner1978}. The step size is chosen adaptively using the so-called \texttt{H211b} digital filter \cite{Soderlind2003, Soderlind2006} to meet the prescribed error tolerance, set as of the order of machine precision. A more detailed study of time-discretization accuracy for similar spectral models can be found in \cite{Klein2011}.


\subsection{Numerical results}

\begin{figure}
  \centering
  \includegraphics[width=0.99\textwidth]{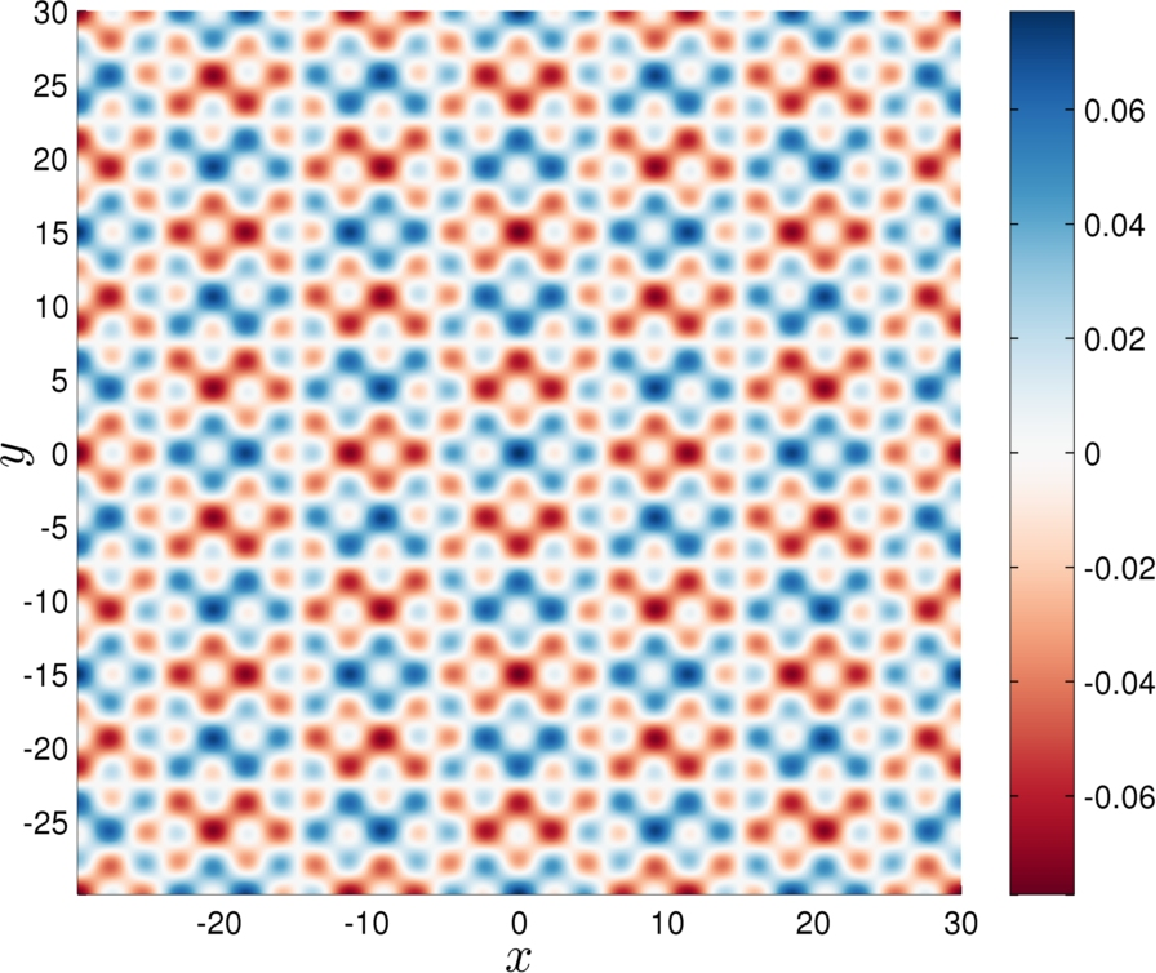}
  \caption{\small\em A bi-periodic wave pattern, $\Omega\ =\ [\,-30,\,30]^{\,2}$ and $\omega\ =\ 0.30\,$.}
  \label{fig:f30}
\end{figure}

\begin{figure}
  \centering
  \subfigure[$t\ =\ 1.60$]{%
  \includegraphics[width=0.49\textwidth]{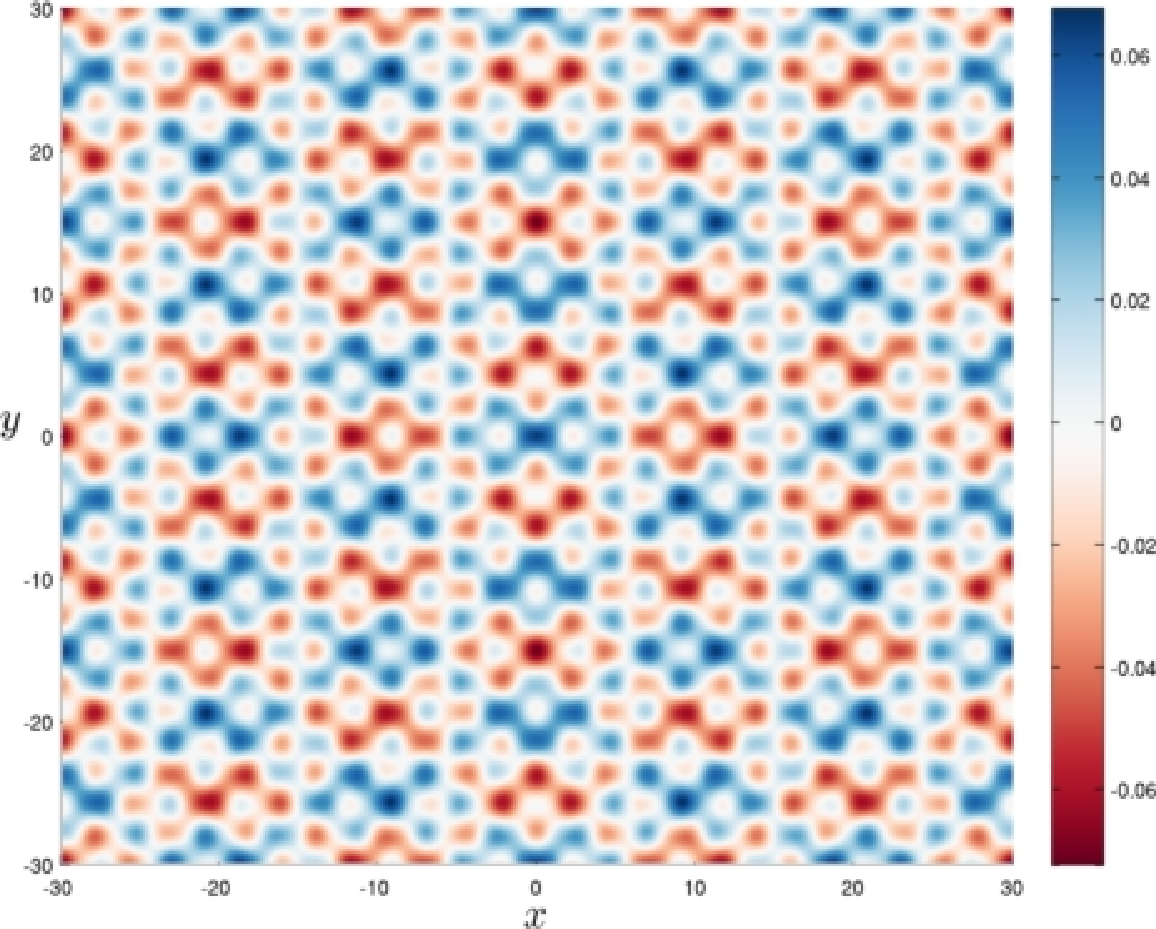}}
  \subfigure[$t\ =\ 8.01$]{%
  \includegraphics[width=0.49\textwidth]{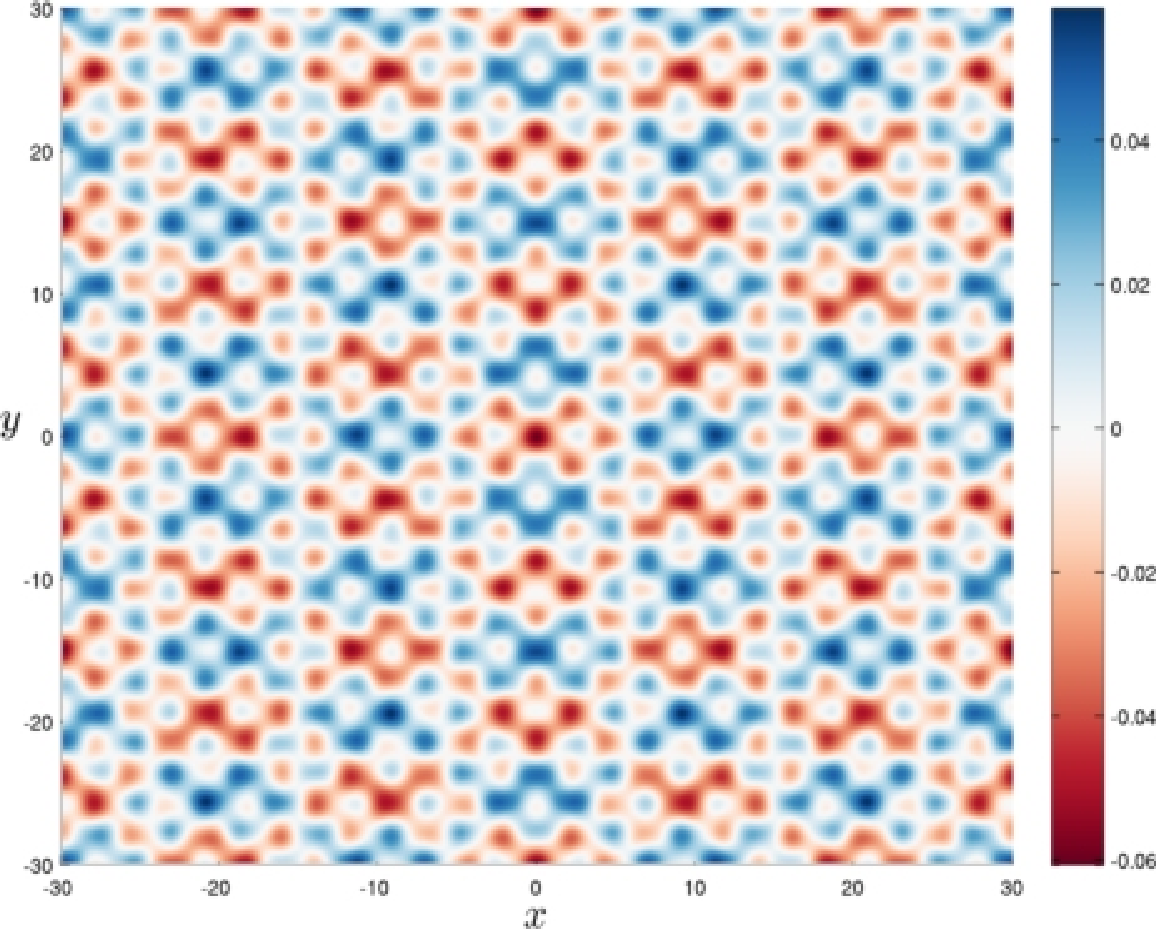}}
  \subfigure[$t\ =\ 11.0$]{%
  \includegraphics[width=0.49\textwidth]{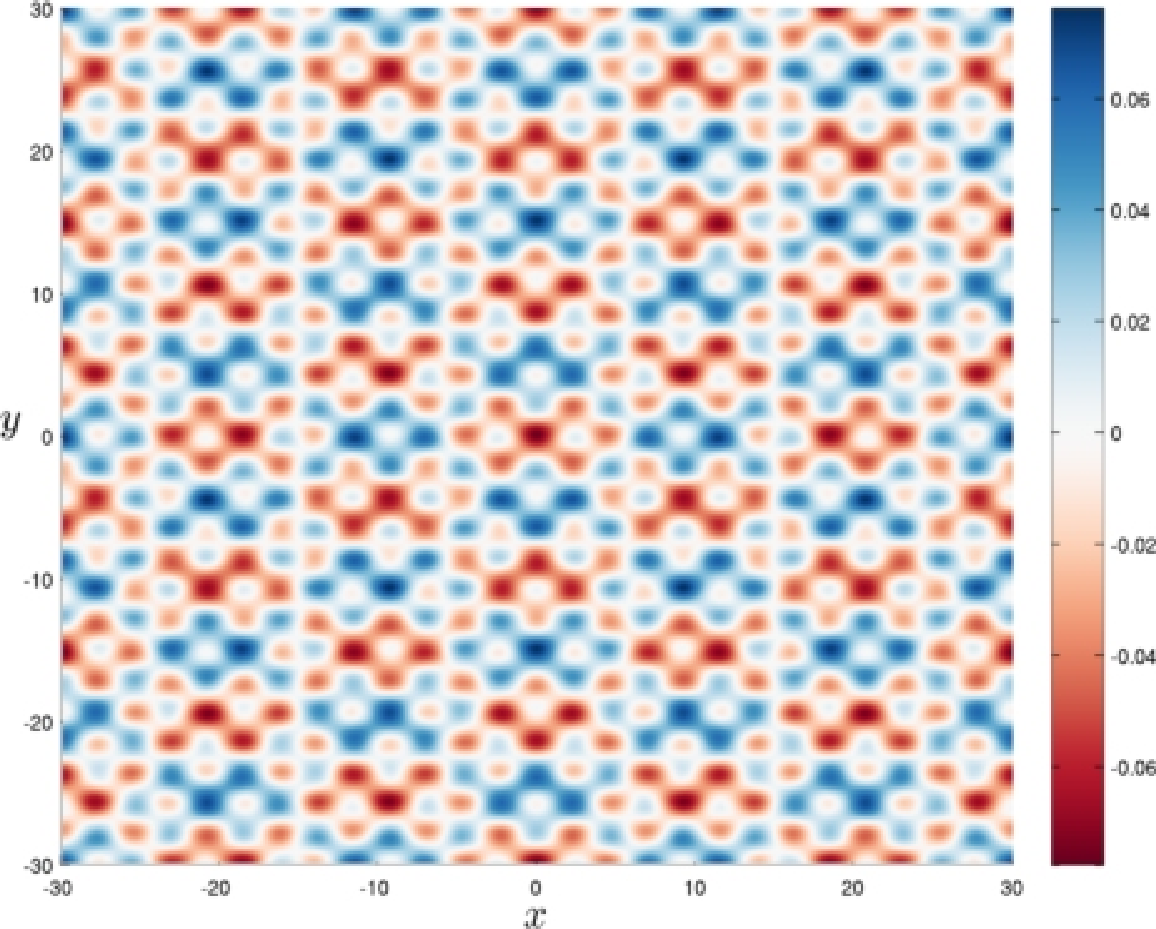}}
  \subfigure[$t\ =\ 14.0$]{%
  \includegraphics[width=0.49\textwidth]{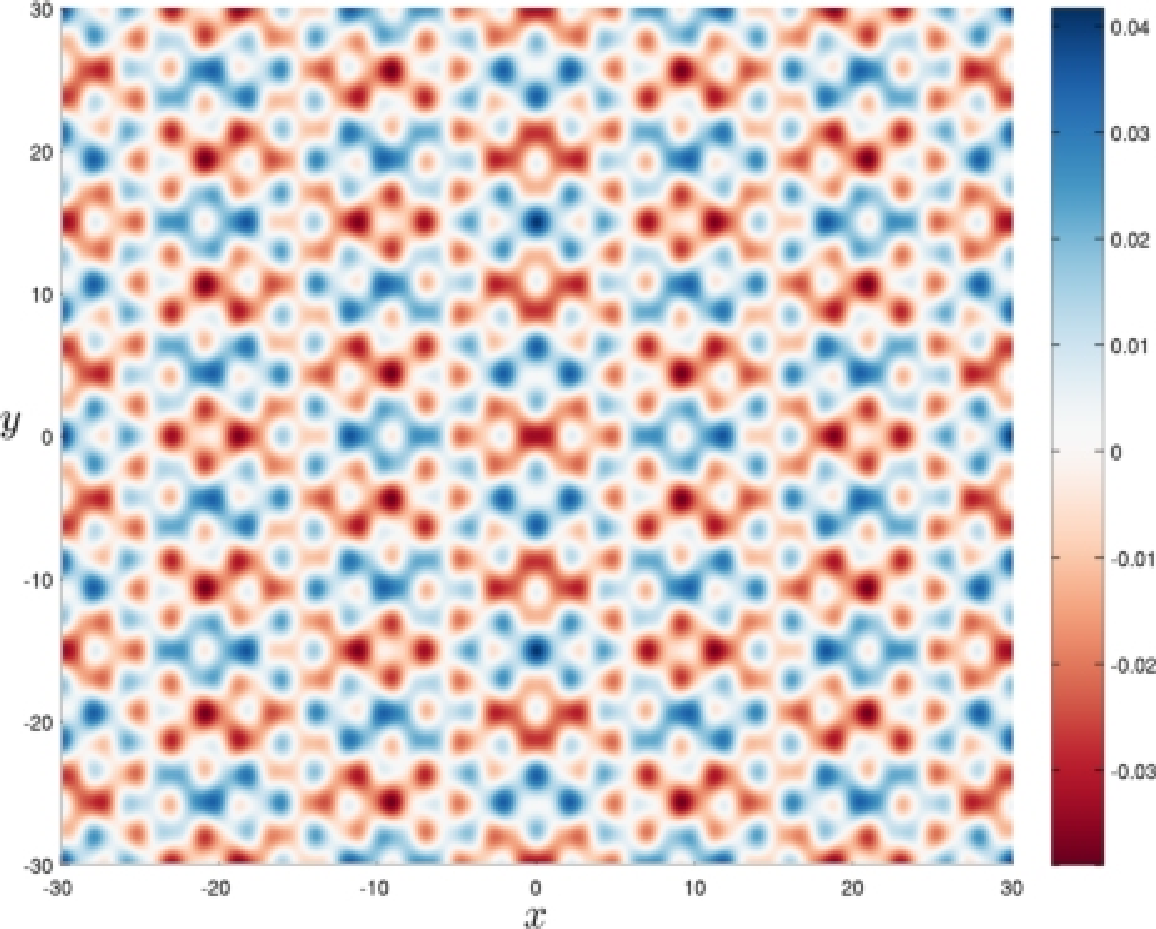}}
  \caption{\small\em Snapshots taken from the dynamic simulation of the periodic pattern represented in Figure~\ref{fig:f30}.}
  \label{fig:dyn}
\end{figure}

Consider the bi-periodic pattern computed for $\Omega\ =\ [\,-30,\, 30\,]^{\,2}$ and $\omega\ =\ 0.30$ and shown in Figure~\ref{fig:f30}. To simulate the dynamics the pseudo-spectral method described above is used with $256\times 256$ \textsc{Fourier} modes. The numerical solver independently confirmed that the time-harmonic standing wave $A\,(x,\,y,\,t)\ =\ B\,(x,\,y)\,\ue^{\,-\ui\,\omega\, t}$ associated to the spatial pattern $B\,(x,\,y)$ of Figure~\ref{fig:f30} computed using the \textsc{Petviashvili} scheme is effectively a solution of the original \acs{hNLS} equation \eqref{eq:adim}. Then, another experiment was performed where the initial condition for the solver was set as $A\,(x,\,y,\,0)\ =\ B\,(x,\,y)\ +\ w\,(x,\,y)\,$, where $w\,(x,\,y)$ is approximatively a $7\%$ double-periodic perturbation with the wavelength four times smaller than that of the unperturbed pattern $B\,(x,\,y)\,$. The simulations were carried out up to the dimensionless time $T\ =\ 14.0\,$. The energy (\textsc{Hamiltonian}) $\Hh$ and action $\Aa$ were conserved with $12$ digits accuracy during the whole simulation. The momentum $\Mm$ was preserved to machine precision. On short time scales slight oscillations around the unperturbed solution occur. However, on a much longer time scale a transition to another solution is observed, which is quite similar in shape to the $B\,(x,\,y)\,$, but slightly shifted in space. A few snapshots taken from the dynamical simulation are depicted in Figure~\ref{fig:dyn} (see also video \cite{Vuillon2014}).

\begin{figure}
  \centering
  \subfigure[]{%
  \includegraphics[width=0.49\textwidth]{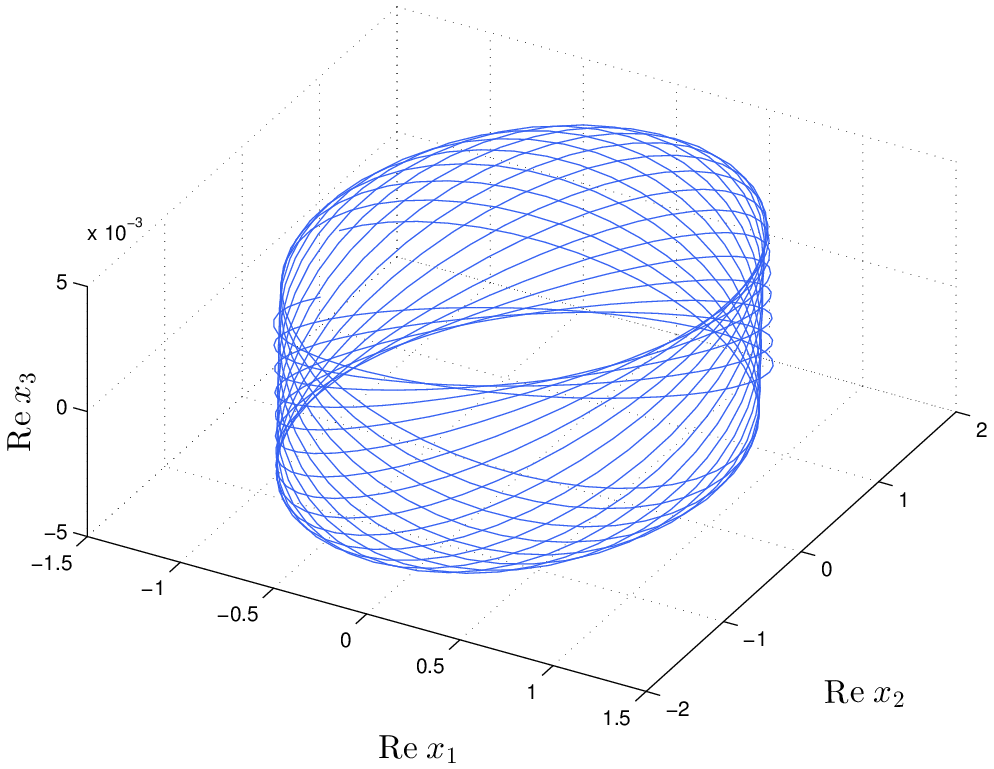}}
  \subfigure[KL Mode 1]{%
  \includegraphics[width=0.49\textwidth]{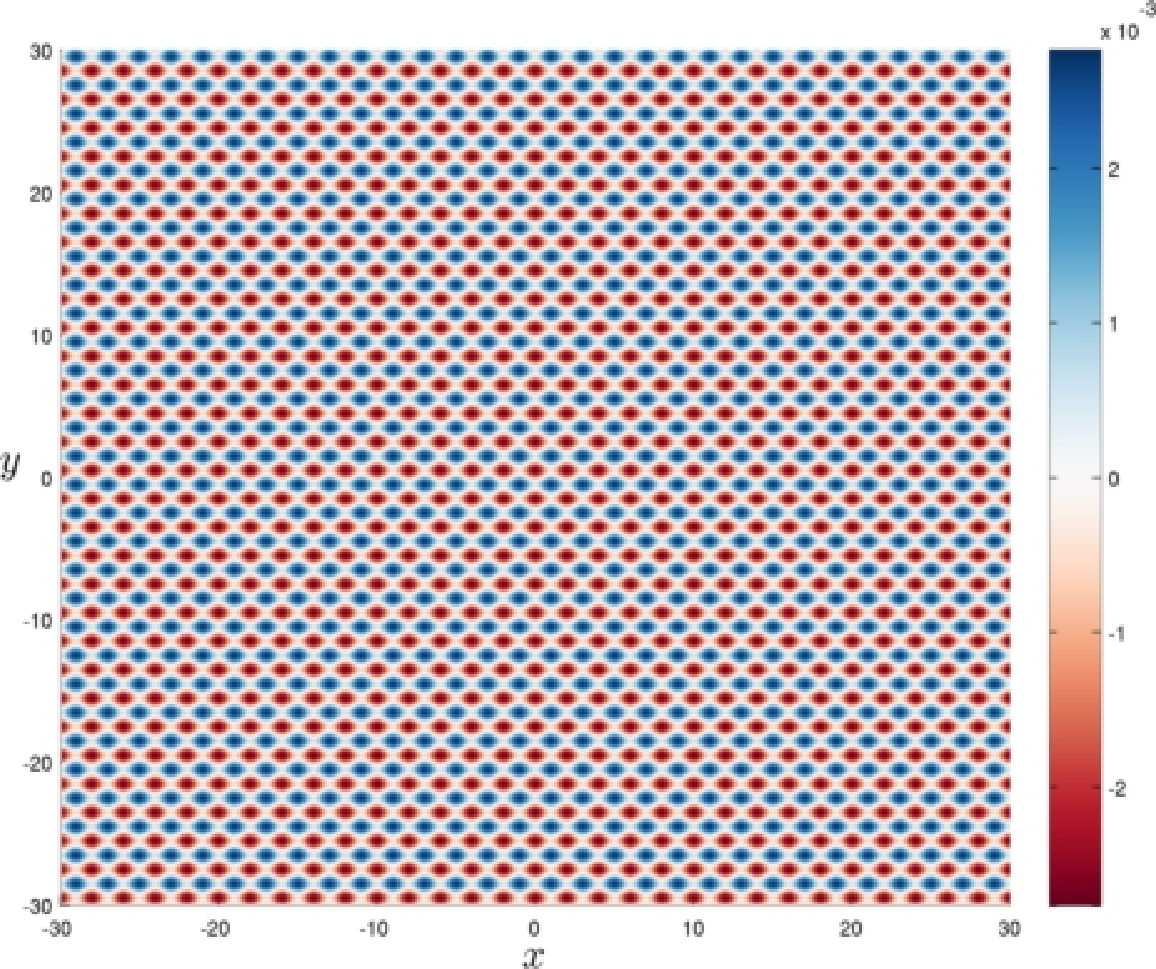}}
  \subfigure[KL Mode 2]{%
  \includegraphics[width=0.49\textwidth]{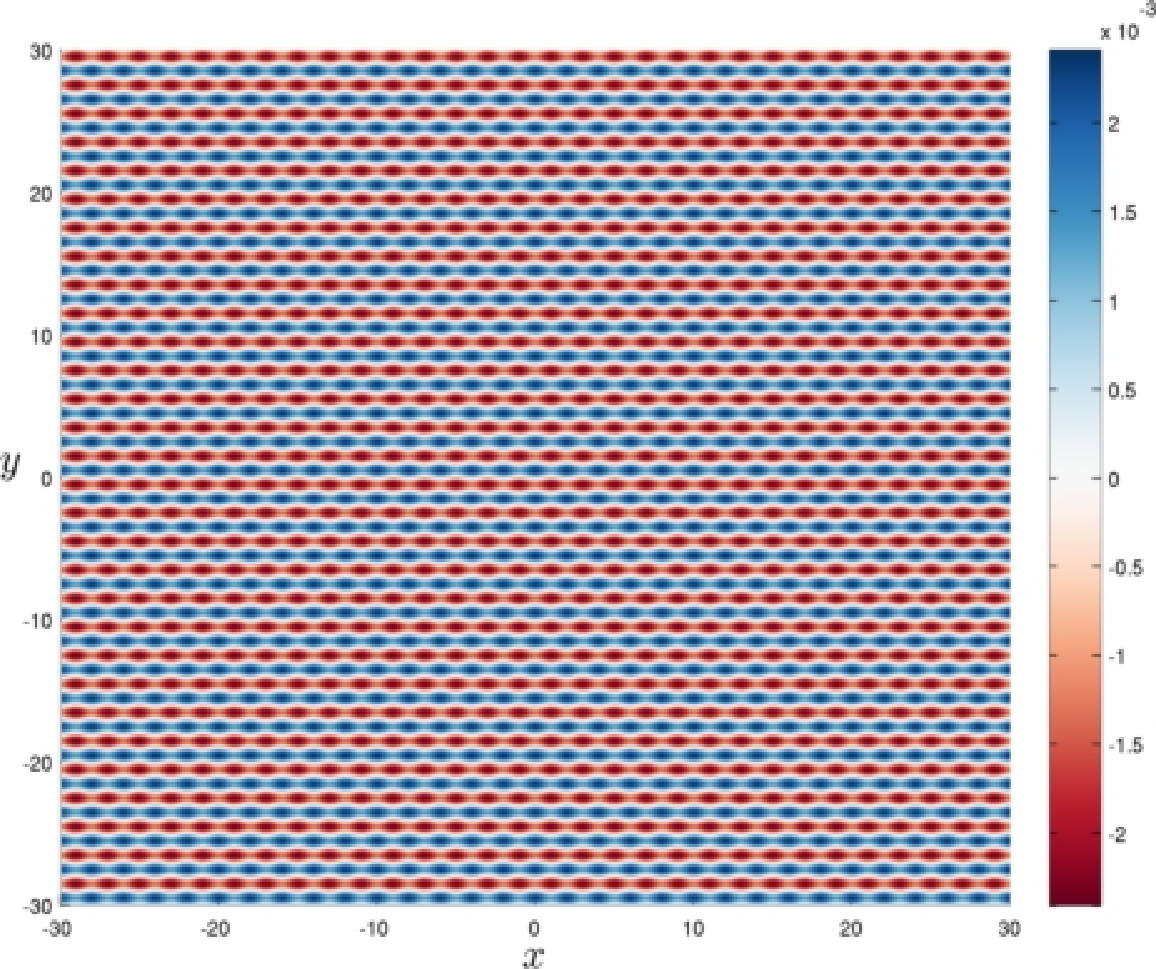}}
  \subfigure[KL Mode 3]{%
  \includegraphics[width=0.49\textwidth]{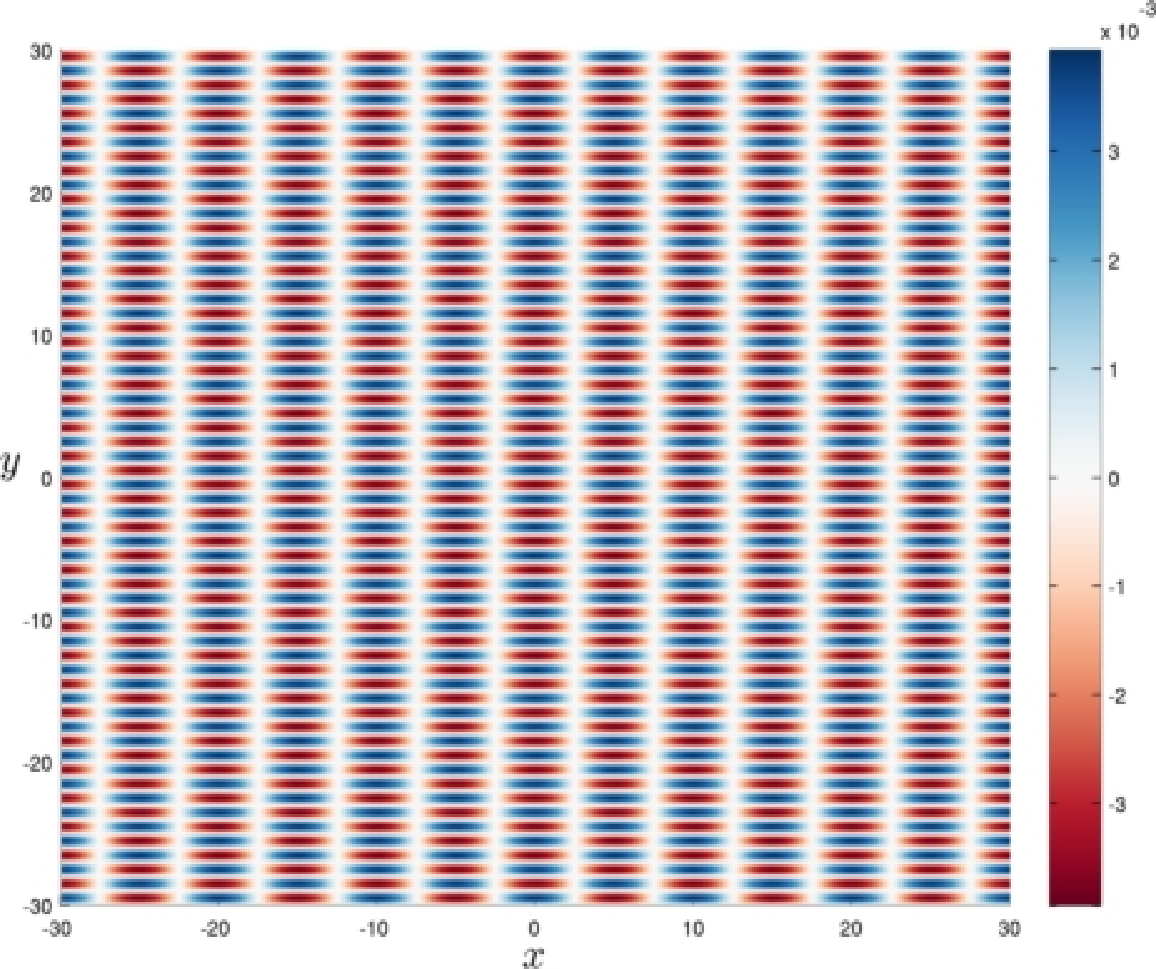}}
  \caption{\small\em (a) Trajectory $\gamma\,(t)$ in the subspace $\S\ =\ \spn\bigl\{\phi_{\,1},\, \phi_{\,2},\, \phi_{\,3}\bigr\}$ spanned by the three dominant KL modes $\phi_{\,j}\,$, $j\ =\ 1,\,\ldots,\,3$ associated to the simulated dynamics of Figure~\ref{fig:f30} (b) first, (c) second and (d) third KL mode.}
  \label{fig:KL}
\end{figure}

To visualize the dynamics, $A$ is projected onto the subspace $\S\ =\ \spn\bigl\{\phi_{\,1},\, \phi_{\,2},\, \phi_{\,3}\bigr\}$  spanned by the first three leading \textsc{Karhunen}--\textsc{Lo\`eve} (KL) eigenmodes  $\phi_{\,j}\,$, $j\ =\ 1,\,2,\,3$ (see, for example \cite{Ghanem2003}). These are estimated from the numerical simulations using the method of snapshots (see \cite{Sirovich1987, Rowley2005}) after the time average is removed. The associated trajectory $\gamma\,(t)$ in $\S$ and the three KL modes are shown in Figure~\ref{fig:KL}. The first two modes represent the most energetic structures of the imposed perturbation, whereas the 3\up{rd} mode arises due to the nonlinear interaction between the unperturbed standing wave and the perturbation. Note that $\gamma$ lies approximately on a cylindrical manifold. The motion is circular on the $x_{\,1}$ -- $x_{\,2}$ plane with oscillations in the vertical $x_{\,3}$ axis. In physical space the dynamical wave patterns smoothly vary between the 1\up{st} and 2\up{nd} KL mode in a periodic fashion while being modulated by the 3\up{rd} mode. A new dynamical state is reached, which is not a standing wave.

\section{Conclusions}
\label{sec:concl}

In this study a class of special solutions to the hyperbolic NLS equation \eqref{eq:hNLS} have been investigated. In particular, bi-periodic standing waves are obtained numerically using the iterative \textsc{Petviashvili} scheme \cite{Petviashvili1976, Pelinovsky2004, Lakoba2007}. Non-trivial wave patterns are revealed by varying the computational domain $\Omega$ and the frequency $\omega$ of the standing wave. These are described by means of symbolic dynamics and the language of substitutions. For given value of $\omega\,$, the patterns associated to different domains $\Omega$ have in common the same substitution rule $\sigma\,$. The dynamics of a perturbed standing wave is also numerically investigated by means a highly accurate \textsc{Fourier} solver in the reduced state space $S$ spanned by the first three dominant KL eigenmodes. The trajectory in $S$ lies approximately on a cylindrical manifold and in physical space the wave pattern appears to vary both in space and time. The discrete symbolic construction is the first step for developing a whole theory of description of periodic patterns of the \acf{hNLS} equations. The theoretical explanation of steady and unsteady wave patterns via coding remains a major challenge and also a perspective opened by this study. The main novelty of our study consists in an attempt to interpret the spatial patterns using the tools from the combinatorics. These ideas are applied to a fundamentally nonlinear and practically important PDE. The nonlinearity is embraced in its full complexity without any simplifying assumptions.


\subsection{Perspectives}

In future studies we would like to apply the symbolic coding technique to other PDEs, which are able to generate spatial $1-$D and $2-$D patterns. The $3-$D case is certainly more ambitious, but it is on our agenda as well. On the other side, we have the stability questions of computed patterns. In the present study we addressed this question by simulating the perturbed patterns and our results indicate the stability. However, the linear spectral stability can be established using the \textsc{Floquet}-type approaches.


\subsection*{Acknowledgments}
\addcontentsline{toc}{subsection}{Acknowledgments}

D.~\textsc{Dutykh} acknowledges the support from ERC under the research project ERC-2011-AdG 290562-MULTIWAVE. The authors would like to thank Professor Angel~\textsc{Duran}\footnote{University of Valladolid, Spain} for helpful discussions on the \textsc{Petviashvili} method. We would like also to thank Pavel~\textsc{Holodoborodko}\footnote{LLC Advanpix, Japan} for providing us with the MCT Toolbox for \textsc{Matlab}. We thank Professor Alexei~\textsc{Cheviakov} (University of Saskatchewan, Canada) for his support in using the \textsc{GeM} package for \textsc{Maple}. Finally, we thank Professor Olivier~\textsc{Le Gal} (University Savoie Mont Blanc, France) for helpful discussions on the geometry of point transformations.


\appendix
\section{Conservation laws of the hyperbolic NLS equation}

The conservation laws of the last system \eqref{eq:adim1}, \eqref{eq:adim2} can be computed using sophisticated computer algebra methods\footnote{We employed the \textsc{Maple} package \textsc{GeM}.} \cite{Cheviakov2007}:
\begin{multline*}
  (1) \quad (u\,v_{\,x})_{\,t}\ -\ \frac{1}{2}\;\Bigl[\,\bigl(u^{\,2}\ +\ v^{\,2}\bigr)^2\ +\ u_{\,x}^{\,2}\ +\ u_{\,y}^{\,2}\ +\ v_{\,x}^{\,2}\ +\ v_{\,y}^{\,2}\,\Bigr]_{\,x}\\
  +\ \Bigl[\,u_{\,x}\,v_{\,y}\ +\ v_{\,x}\,v_{\,y}\,\Bigr]_{\,y}\ =\ 0\,,
\end{multline*}
\begin{multline*}
  (2) \quad (u\,v_{\,y})_{\,t}\ -\ \Bigl[\,u_{\,x}\,u_{\,y}\ +\ v_{\,x}\,v_{\,y}\ +\ 2\,x\,v^{\,3}\,v_{\,y}\,\Bigr]_{\,x}\ +\\ \frac{1}{2}\;\Bigl[\,u_{\,x}^{\,2}\ +\ u_{\,y}^{\,2}\ +\ v_{\,x}^{\,2}\ +\ v_{\,y}^{\,2}\ -\ 2\,u^{\,2}\,v^{\,2}\ -\ 2\,u\,v_{\,t}\ -\ u^{\,4}\ +\ 4\,x\,v_{\,x}\,v^{\,3}\,\Bigr]_{\,y}\ =\ 0\,,
\end{multline*}
\begin{multline*}
  (3) \quad \bigl(x\,u\,v_{\,y}\ +\ y\,u\,v_{\,x}\bigr)_{\,t}\ -\ \frac{1}{2}\;\Bigl[\,y\,\bigl((u^{\,2}\ +\ v^{\,2})^{\,2}\ +\ u_{\,x}^{\,2}\ +\ u_{\,y}^{\,2}\ +\ v_{\,x}^{\,2}\ +\ v_{\,y}^{\,2}\ +\ 2\,u\,v_{\,t}\bigr)\ +\\ 2\,x\,\bigl(u_{\,x}\,u_{\,y}\ +\ v_{\,x}\,v_{\,y}\ +\ x\,v^{\,3}\,v_{\,y}\bigr)\,\Bigr]_{\,x}\ + \\
  \frac{1}{2}\;\Bigl[\,x\,\bigl(u_{\,x}^{\,2}\ +\ u_{\,y}^{\,2}\ +\ v_{\,x}^{\,2}\ +\ v_{\,y}^{\,2}\ -\ (u^{\,2}\ +\ v^{\,2})^{\,2}\ +\ v^{\,4}\ -\ 2\,u\,v_{\,t}\ +\ 2\,x\,v^{\,3}\,v_{\,x}\bigr)\\ +\ 2\,y\,\bigl(u_{\,x}\,u_{\,y}\ +\ v_{\,x}\,v_{\,y}\bigr)\,\Bigr]_{\,y}\ =\ 0\,,
\end{multline*}
\begin{equation*}
  (4) \quad \Bigl(\frac{1}{2}\;\bigl(u^{\,2}\ +\ v^{\,2}\bigr)\Bigr)_{\,t}\ +\ \Bigl[\,u_{\,x}\,v\ -\ u\,v_{\,x}\,\Bigr]_{\,x}\ +\ \Bigl[\,u\,v_{\,y}\ -\ v\,u_{\,y}\,\Bigr]_{\,y}\ =\ 0\,,
\end{equation*}
\begin{multline*}
  (5) \quad \Bigl(\frac{1}{2}\;x\,\bigl(u^{\,2}\ +\ v^{\,2}\bigr)\ +\ 2\,t\,u\,v_{\,x}\Bigr)_{\,t}\ -\ \Bigl[\,t\,\bigl((u^{\,2}\ +\ v^{\,2})^{\,2}\ +\ u_{\,x}^{\,2}\ +\ u_{\,y}^{\,2}\ +\ v_{\,x}^{\,2}\ +\ v_{\,y}^{\,2}\ +\ 2\,u\,v_{\,t}\bigr)\ +\\
  x\,\bigl(u\,v_{\,x}\ -\ v\,u_{\,x}\bigr)\ +\ u\,v\,\Bigr]_{\,x}\ +\\
  \Bigl[\,2\,t\,\bigl(u_{\,x}\,u_{\,y}\ +\ v_{\,x}\,v_{\,y}\bigr)\ +\ x\,\bigl(u\,v_{\,y}\ -\ v\,u_{\,y}\bigr)\,\Bigr]_{\,y}\ =\ 0\,,
\end{multline*}
\begin{multline*}
  (6) \quad \Bigl(\frac{1}{2}\;y\,\bigl(u^{\,2}\ +\ v^{\,2}\bigr)\ -\ 2\,t\,u\,v_{\,y}\Bigr)_{\,t}\ +\ \Bigl[\,2\,t\,\bigl(u_{\,x}\,u_{\,y}\ +\ v_{\,x}\,v_{\,y}\ +\ 2\,x\,v^{\,3}\,v_{\,y}\bigr)\ +\ y\,\bigl(v\,u_{\,x}\ -\ u\,v_{\,x}\bigr)\,\Bigr]_{\,x}\\
  -\ \Bigl[\,t\,\bigl((u^{\,2}\ +\ v^{\,2})^{\,2}\ -\ v^{\,4}\ +\ 2\,u\,v_{\,t}\ +\ u_{\,x}^{\,2}\ +\ u_{\,y}^{\,2}\ +\ v_{\,x}^{\,2}\ +\ v_{\,y}^{\,2}\ -\ 2\,x\,v^{\,3}\,v_{\,x}\bigr)\\
  +\ y\,\bigl(u\,v_{\,y}\ -\ v\,u_{\,y}\bigr)\ +\ u\,v\,\Bigr]_{\,y}\ =\ 0\,.
\end{multline*}
The last relations can be used in theoretical studies, but also to check the accuracy of numerical computations.



\bibliography{biblio}
\bigskip

\end{document}